\documentclass{aa}

\usepackage[latin1]{inputenc}
\usepackage{moreverb}
\usepackage{fancyhdr}
\usepackage{aalongtable}
\usepackage{rotating}
\usepackage{graphicx}
\usepackage{amsmath}
\usepackage{amssymb}

\usepackage{natbib}
\bibpunct{(}{)}{;}{a}{}{,}

\usepackage[english]{babel}

\defcitealias{k2003}{K2003}
\defcitealias{g2003}{G2003}
\defcitealias{v2001}{V2001}

\def\vone{\mbox{v=1--0~S(1)}}

\def\kms{~\mbox{km\,s$^{-1}$}}
\def\cm{~\mbox{cm$^{-3}$}}
\def\wmsr{~\mbox{W\,m$^{-2}$\,sr$^{-1}$}} 
\def\thetc{$\theta^1$Ori~C}
\def\hii{\ion{H}{ii}}
\newcommand{\refe}[1]{{#1}}

\begin{document}

\title{Observations of spatial and velocity structure in the Orion Molecular
  Cloud\thanks{Based on observations performed at the CFHT 3.6~m
    telescope}}

\author{H.~D.~Nissen\inst{1}
\and M.~Gustafsson\inst{1}
\and J.L.~Lemaire\inst{2}\fnmsep $^{\star \star}$
\and Y.~Cl\'enet\inst{3}
\and D.~Rouan\inst{3}
\and D.~Field\inst{1}\fnmsep\thanks{Visiting astronomer at the
  Canada-France-Hawaii Telescope, Mauna Kea, Hawaii}
}

\institute{Department of Physics and Astronomy, University of Aarhus,
  8000 Aarhus C, Denmark 
\and Observatoire de Paris and Universit\'e de Cergy-Pontoise, LERMA and
UMR 8112 du CNRS, 92195 Meudon, France
\and Observatoire de Paris-Meudon, LESIA, 92195 Meudon Principal Cedex, France
}

\offprints{D. Field, \email{dfield@phys.au.dk}}

\date{Received \today / Accepted \today}

\abstract{Observations are reported of H$_2$ IR emission in the S(1)
  v=1-0 line at 2.121~$\mu$m in the Orion Molecular Cloud, OMC1, using
  the GriF instrument on the Canada-France-Hawaii Telescope. GriF uses a
  combination of adaptive optics and Fabry-Perot interferometry,
  yielding a spatial resolution of 0.15\arcsec\ to 0.18\arcsec\ and a

  velocity discrimination as high as 1\kms. Thanks to the high spatial
  and velocity resolution of the GriF data, 193 bright H$_2$ emission
  regions can be identified in OMC1. The general characteristics of
  these features are described in terms of radial velocities,
  brightness and spatial displacement of maxima of velocity and
  brightness, the latter to yield the orientation of flows in the
  plane of the sky. Strong spatial correlation between velocity and
  bright H$_2$ emission is found and serves to identify many features
  as shocks. Important results are: \emph{(i)} velocities of the
  excited gas illustrate the presence of a zone to the south of
  BN-IRc2 and Peak 1, and the west  of Peak 2, where there is a
  powerful blue-shifted outflow with an average velocity of
  $-$18\kms. This is shown to be the NIR counterpart of an outflow
  identified in the radio from source I, a very young
  O-star. \emph{(ii)} There is a band of weak velocity features
  ($<$5\kms) in Peak 1 which may share a common origin through an
  explosive event, in the BN-IRc2 region, with the fast-moving
  fingers (or bullets) to the NW of OMC1. \emph{(iii)} A proportion of
  the flows are likely to represent sites of low mass star formation
  and several regions show multiple outflows, probably indicative of
  multiple star formation within OMC1. The high spatial and velocity
  resolution of the GriF data show these and other features in more
  detail than has previously been possible.

}

\titlerunning{Shocks in Orion}

\maketitle


\section{Introduction}

The Orion Molecular Cloud (OMC1) \citep[D\:=460~pc,][]{bally00} is the
closest highly active massive star-forming region. It has
accordingly been studied over a large range of wavelengths from the
radio to the X-ray \citep[e.g. ][]{genzel89,wright96,odell01,chrysostomou00,
  garmire00,ferland01,feigelson02,doi02, odell03,tan04} and has become the
type-site for studies of star formation. Many features are accordingly
very well-described, for example the detailed morphology of the HII
region associated with \thetc, the dominant source of ionizing
radiation in the Trapezium \citep{odell01}. Perhaps the principal
significance of OMC1 and its vicinity is that it has been shown to
harbour upward of 1000 young low mass stars in
the presence of a number of high mass stars \citep{hillenbrand97}. 

\begin{figure}
\resizebox{\hsize}{!}{\includegraphics{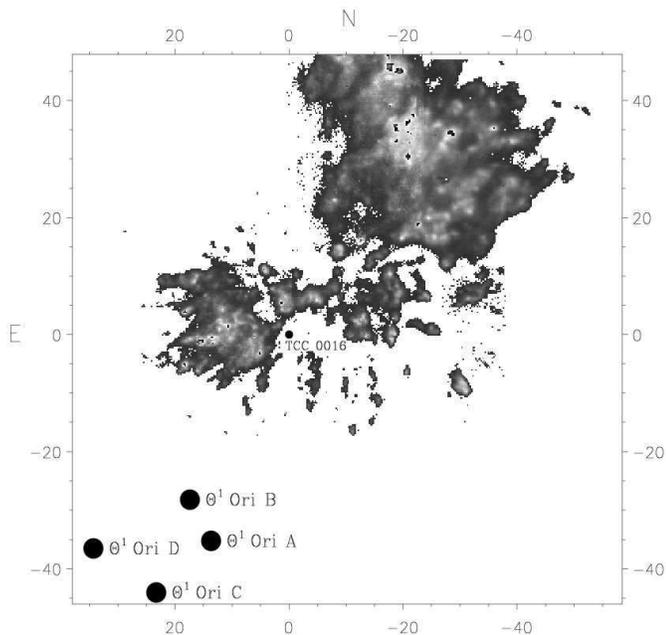}}
\caption{The observed area relative to the Trapezium stars (large
  black dots). The grey scale image show emission brightness in the
  H$_2$ \vone line. Axes are labelled in arcseconds. The star TCC0016,
  at coordinates 0\arcsec~E, 0\arcsec~N and marked with a small black
  dot, is used as a reference for coordinates throughout this work.}
\label{where}
\end{figure}

\begin{figure}
\resizebox{\hsize}{!}{\includegraphics{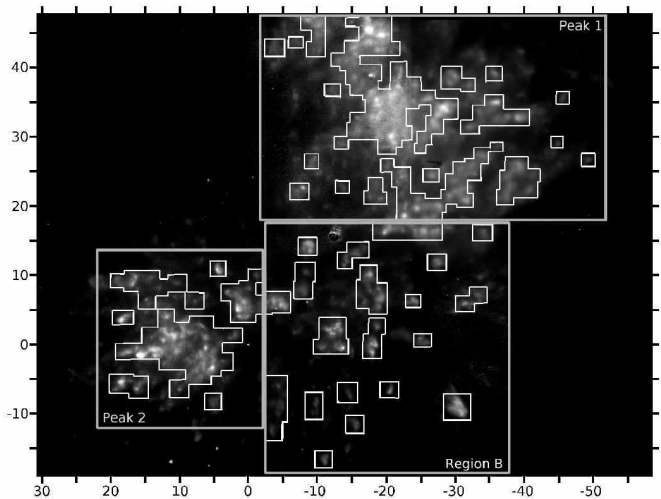}}
\caption{The regions analysed in this work outlined as small
  overlapping white boxes. The gray-scale shows emission in the
  H$_{2}$~v=1--0~S(1) emission line. Also shown as large white boxes
  is the naming convention for Peak~1, Peak~2 \citep{beckwith78}, and
  Region B, the latter bounded by
  \mbox{$-2$\arcsec\,--\,$-40$\arcsec~E},
  \mbox{$-18$\arcsec\,--\,$17$\arcsec~N}, relative to TCC0016. Axes
  are labelled in arcseconds.} 
\label{areamap}
\end{figure}

The present work, covering the area shown in Figs.~\ref{where} and
\ref{areamap},  concentrates upon infrared observations of 
molecular hydrogen (H$_2$), and in particular upon data displaying the
dynamics of OMC1. The gas in OMC1 is highly structured with many bright knots 
and there is ample evidence of much violent
motion within OMC1 over a range of scales from 70~AU to 3 x 10$^4$ AU
\citep{g2006a}. Earlier work concerning gas dynamics
\citep{chrysostomou97,salas99,g2003} showed the presence of numerous
gas flows with flow velocities varying from a few\kms\ in relatively
dense gas to several hundred\kms\ \citep{lee00,doi02}. \refe{The latter are
associated with perhaps the most prominent feature in OMC1, the H$_2$
''fingers'' (also known as ''bullets'') first observed by
\citet{allen93}. The fingers lie principally to the NW of OMC1 and
they are the wake of fast-moving, diffuse clumps of gas  moving
radially out from OMC1 \citep{burton97,lee00,doi02}. 
We do not find any high velocity fingers in the observed field. This
may be due to the limitations of the observations which confine us to
velocities below about 150\kms\ but may also be due to the technique
of data reduction, as further described in sec.~\ref{sec:datared}. 
}

\refe{Two main categories of models have been proposed to explain the
overall structure of OMC1. One is that structures in OMC1 are caused
by shock instabilities which form the bright knots and bullets as 
in \citet{stone95} and \citet{mccaughrean97}. In this model
Rayleigh-Taylor instabilities form in the wind from the central massive
star (or stars) in OMC1 when a faster wind overtakes the dense shell of
material swept up by an earlier, slower wind. This results in a
fragmentation of the shell into smaller knots some of which appears as
bullets, forming fingers of emission, when moving outwards from the
shell at high speed. The other group of models
\citep[e.g.][]{allen93,doi02} explain the bright knots as internal
shocks in the cloud. These can have many origins: turbulent bipolar 
outflows from low-mass star formation or some form of outflow from the
central massive stars (BN, source~I, source~n
\citep{menten95,greenhill04a,shuping04}). The fingers and bullets
themselves in either model are typically interpreted as the result of
some major explosive event involving one or more of the massive
stars. The fingers may then have shot through Peak~1 rather than being
formed there. The lack of fingers and bullets to the SE of OMC1 could
suggest a highly directional origin, but may equally well be explained
by the emitting H$_2$ being too deeply buried to be seen, or it has
been evaporated by the action of the Trapezium stars; if you placed
the bullets to the NW in the corresponding position to the SE, they
would have come very close to the Trapezium. For further discussion of
the fingers and bullets we refer the reader to for example
\citet{axon84,allen93,burton97,odell97c,lee00,odell01,doi02}.}

\refe{In this work we would favour the second group of models mentioned above,
in which a significant proportion of the excitation is internal to
Peak~1 (see fig.~\ref{areamap} for nomenclature). We also note there
are no prominent fingers in Peak 2 and therefore Peak 2 must
presumably have some internal excitation source. In separate work we
have also studied the global nature of the gas motions in OMC1
\citep{g2006a}. In that work a statistical analysis is presented of
the turbulent velocity structure in the Orion Molecular Cloud at
scales ranging from 70~AU to $3 \times 10^4$~AU, using the data of the
present paper. The variance for the velocity field shows only an
approximate power law dependence on lag, with deviations below 2000 AU
which are attributed to outflows and possibly disk structures, below
300~AU, associated with low mass star formation within OMC1. The major
point is that the analysis in \cite{g2006a} demonstrates the presence
of one or more preferred scales in the excited gas. These preferred
scales are most naturally attributed to star formation where
deviations from power law behaviour is associated with injection of
energy at and below the star-forming scale.}

Energy injection into OMC1 is evidently triggered within the molecular
cloud at different scales. Massive stars such as the BN object, a
B-star \citep{gezari98}, or radio source I (hereafter simply source~I), a
deeply buried O-star \citep{menten95,greenhill04a,shuping04}, are
associated with structure at a scale of $\sim$0.1~pc (20,000~AU). Low
mass star formation gives rise to flows and associated structure at
scales of hundreds to several thousand AU. 

The region studied in the present work is shown in Fig. 1 and lies NW
of the Trapezium stars. The target area covers only the inner part of
OMC1, and as we have noted largely excludes for example bullets and
fingers. Thus the material in the present study encompasses 
the gas volume that tends to be subject to denser, slower outflows,
associated with both massive star formation and low mass star
formation.

High spatial resolution K-band (2-2.5~$\mu$m) images of IR emission
from H$_2$ in OMC1 using the HST \citep{stolovy98,chen98,schultz99,doi02},
the Canada-France-Hawaii Telescope (CFHT) \citep{g2003}, the ESO 3.6m
\citep{v2001,k2003}, the VLT \citep{lacombe04} and other telescopes
\citep[e.g.][]{mccaughrean97,schild97} have provided the most recent
evidence for the presence of fast flows and shocks at small scales in
OMC1. 


The recent paper of \citet{g2003}, hereafter Paper I, provided an initial
report of H$_2$ emission data for OMC1 obtained using the so-called
''GriF'' instrument \citep{clenet02} on the CFHT in December
2000. This instrument combines Fabry-Perot interferometry with the
PUEO adaptive optics system on the CFHT \citep{rigaut98}. These data,
containing information on both morphology and velocity, showed
striking evidence for the presence of shocks within OMC1. This was
provided through the clear association of gas motions of tens of\kms\
with bright IR emission of H$_2$. The novelty of these data is
contained in the association of a very high spatial resolution  of
0.15\arcsec\ (70~AU) to 0.18\arcsec\ with velocity discrimination as
good as 1\kms\ (3$\sigma$) in regions of high emission brightness,
obtained with a Fabry-Perot of relatively low resolution of
150\kms. These figures may be compared with the high inherent spectral
resolution of 14\kms\ in \citet{chrysostomou97} and 24\kms\ in
\citet{salas99}, where data were recorded with a spatial resolution
which lay between 1.5\arcsec\ and 2.0\arcsec.
%
%
%
%
%
%
%
In the present work the GriF data are reported in detail. Whereas in
Paper I 19 regions were identified for study, 193 regions are included
here. This affords a much broader perspective on the nature of flows
within OMC1 and represents a detailed attempt at extracting the wealth
of information contained within the GriF observations. New features
are revealed through this extended study of the GriF 
results. In particular, a massive and very energetic blue-shifted
outflow is identified in Region B between Peak~1 and Peak~2
(fig.~\ref{areamap}). \citet{chrysostomou97}, working in K-band at a
spatial resolution of 1.5\arcsec\ found evidence of clumps of gas
dominated by blue-shifted motions in the same zone. As described in
sect.~\ref{sec:d-outflow}, there is a considerable independent body of
evidence from radio data for an outflow in this zone 
\citep{genzel81,genzel89,wright96,greenhill04a,doeleman04}. The
outflow detected here in excited H$_2$ is characterised in some detail
and it is shown that the outflow is the IR signature of the outflow
identified through radio observations. 

On a smaller scale, numerous regions of bright H$_2$ emission are identified
in Peak 1, the NW region, which show only weak radial velocity
components with no clear velocity structure. In addition there are
several regions on the 1000~AU scale with multiple outflows. If, as we
discuss in some detail here, there are protostars associated with some
of these latter zones, then our data present a picture of star
formation that is more complex than the simple disk-bipolar outflow
model of single star formation.

In sect.~\ref{sec:datared} the observations and data reduction,
described in detail in Paper I, are briefly
reviewed. Sect.~\ref{sec:dataanalysis} illustrates and collates the 
data for all 193 regions in terms of radial velocities of flows,
brightness of emission associated with these flows and the
displacement in the plane of the sky between the maximum in emission 
brightness and the maximum in local radial velocity. The mechanism of
H$_2$ excitation is also briefly discussed. In sect.~\ref{sec:results},
the data are presented in order to give a global view of the spatial
distribution of radial flow velocities and the characteristics of
various zones. Among major points that emerge are \emph{(i)} the presence of
the outflow mentioned above between Peaks 1 and 2, in Region~B
(Fig.~\ref{areamap}), \emph{(ii)} the accumulation of zones with high 
H$_2$ brightness but with $<$5\kms\ radial motion in a region passing
diagonally across Peak 1, \emph{(iii)} the greater tendency for shocked
regions to group more closely in Peak 2 than in Peak 1 and \emph{(iv)} the
fact that maxima in brightness of emission and in velocity are in
general slightly displaced. In
sect.~\ref{sec:discuss} a short qualitative account of the structure of
shocks is followed by the use of data to establish the orientations of
shocks within the plane of the sky. Then follows a consideration of
possible sites of current low mass star formation in OMC1. In
sect.~\ref{sec:d-outflow} the origin of the outflow in Region~B and its

spatial association with continuum sources in the IR and radio is
discussed in detail.
%
%
\section{Observations and data reduction} \label{sec:datared}
Observations were performed of the regions shown in Figs.~\ref{where} 
and~\ref{areamap} on December 5${\rm th}$ 2000 at the 3.6m CFHT,
using the then new instrument GriF \citep{clenet02}. This instrument
combines the high spatial resolution of the PUEO adaptive optics with
spectral resolution provided by a Queensgate ET50WF Fabry-Perot
interferometer (FP), giving a 
data-cube showing brightness in each position (pixel) for a series of
wavelengths. The total observed region in OMC1 is a
$\sim$\:1\arcmin\,$\times$\,1\arcmin\ area centered approximately on the
\object{BN}-object, as shown in Fig.~\ref{areamap}.

Observations were made in the K-band using the H$_{2}$~v=1--0~S(1)
emission line, which has a rest wavelength of 2.1212544~$\mu$m
\citep{bragg82}. The line was scanned using the FP with a step size of
$\sim$4.5\,$\times$\,10$^{-4}$~$\mu$m to obtain images showing the
brightness at a series of wavelengths on both sides of the
rest wavelength. A H$_2$ \vone\ interference filter of central wavelength
2.122~$\mu$m and bandwidth of 0.02~$\mu$m was inserted between the FP
and the detector to prevent the superposition of different FP orders. 
The spatial resolution of the data was 0.15\arcsec, estimated from the
PSF of stars in the field. The pixel scale was 0.035\arcsec. The
inherent spectral resolution of the FP was 150\kms. To reduce the 
level of noise, the data have been smoothed with a moving boxcar
average over 3\,$\times$\,3 pixels, resulting in a 20\% loss of
spatial resolution, that is, the final resolution of the data is
0.18\arcsec. For a given position in the plane of the sky the
data give the count rate for each observed wavelength. This is then
fitted with a lorentzian \citep[ and Paper I]{clenet02,chrysostomou97}
to find the maximum count rate and the corresponding  wavelength. This
gives the velocity of the gas at that position. 
Due to high signal-to-noise very precise fits can be made. In bright
regions the uncertainty in the fitted velocity is $\pm$1\kms\ or less
at 3$\sigma$. In the dimmest regions examined in this work (with a
very few exceptions) the uncertainty in the peak in the lorentzian, used to
determine flow velocities (see sect.~\ref{sec:findflows}), is
$\leq$2\kms\ while velocities of the gas surrounding regions of flow
typically have uncertainties of 3-4\kms. A detailed account of the
errors associated with velocity determination may be found in
\citet{g2006a} \refe{where the uncertainty associated with velocity
  determination and the counts per pixel per exposure are explicitly
  related through an empirical formula}.

\refe{Where weaker high velocity emission is seen
spatially coincident with stronger low velocity emission, we record
only the velocity of the brighter slower gas. This is however a rare
occurence and we are aware of only two cases where this has been
encountered: (i) the emission feature 
around position $-$24.2\arcsec~E 6.5\arcsec~N where \citet{stolovy98}
find a weak high velocity component together with the stronger low
velocity feature which we find and (ii) at $-$19.7\arcsec~E
16.2\arcsec~N where we observe the object HH208 at a radial velocity
of $-$19\kms. HH208 has however been identified as a bullet, on the
basis of associated FeII emission, with a radial velocity between
$-$120 and $-$180\kms, \citet{axon84,odell97c}.}

Velocity calibration was carried out by assigning to the mean velocity
of all data a value of 12$\pm$6\kms\ in the local standard of
rest (v$_{lsr}$), consistent with data in \citet{chrysostomou97,
 salas99} and \citet{odell01}. Velocity calibration is however not
material to this work, as we are concerned with relative velocities
over small distances of typically a few arcseconds. For further details
on observations and data reduction see Paper I and references therein. 

Throughout this paper all positions are given in arcseconds east and
north relative to the star TCC0016, which itself is situated at
05$^{\rm h}$35$^{\rm m}$14\fs91, $-$05\degr22\arcmin39\farcs31 (J2000).
%
%
\section{Data analysis} \label{sec:dataanalysis}
\subsection{Identification and characterization of emitting zones}
\label{sec:findflows} 
Data pertaining to the 193 regions analyzed here, are collated in
tables~\ref{table_p11}~--~\ref{table_cz}. All flows are designated by
number in these tables. The tables each refer to one of the three
regions into which the full region has been subdivided, namely Peak~1,
Peak~2 and Region B.

Features were identified on the basis of their brightness and our
concomitant ability to locate region boundaries. Thus all regions show
a well-defined localized increase in brightness compared to the
surrounding gas. Each of these regions has been examined for evidence
of a local gas flow. This was done by making 3D-images of each region,
where these images show radial velocity and emission brightness as a
function of the position in the plane of the sky.
%
\begin{figure*}
\resizebox{\hsize}{!}{\includegraphics{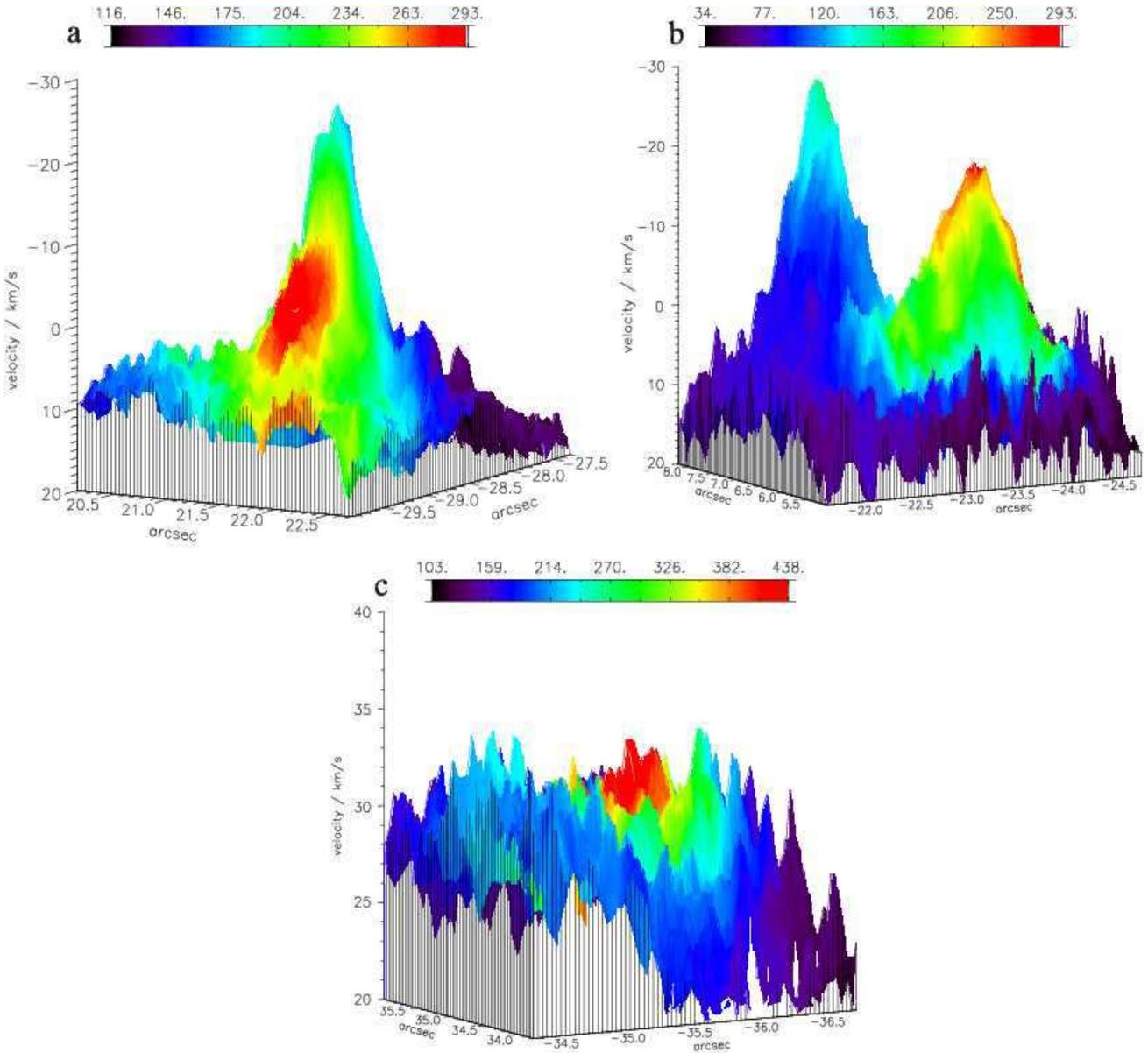}}
\caption{\textbf{a} Flow no. 1-61: see table~\ref{table_p11}. The
  vertical axis shows v$_{lsr}$ in\kms. Colours denote brightness as
  shown on the colour bar above, in counts per 400 sec. The plane of
  the sky is shown in arcseconds, relative to TCC0016. \textbf{b}
  Flows no. B-44, B-45 and B-46: see table~\ref{table_cz}. Otherwise
  as in \textbf{a}. \textbf{c} Flow no. 1-74: see
  table~\ref{table_p11}. This is an example of a flow which shows no
  clear velocity peak exceeding 5\kms.}
\label{3d-demo}
\end{figure*}
%
%
Two illustrative examples of regions exhibiting flows are shown in
Figs.~\ref{3d-demo}a and b, where the x- and y-axes give
the position in the plane of the sky and the z-axis shows the radial
velocity of the gas. A colour code associated with every image shows
the observed brightness as the number of counts per exposure of
400~s. This can be converted to brightness through the use of the value of
3.0$\pm$0.15\,$\times$\,10$^{-5}$\wmsr~found in \citet{v2001} for the
brightest region observed here.\footnote{The brightest region is in
  Peak~2 at coordinates 15.82\arcsec\ E, $-$1.61\arcsec\ N, see
  table~\ref{table_p2}, flow no. 2-10.} Using this value it is found that
a count of 100 over the 400~s exposure corresponds to
$\sim$3.5\,$\times$\,10$^{-6}$\wmsr. This conversion may be applied to 
all regions (and all figures in this work) with the proviso that
systematic errors may arise through differing conditions of airmass
and of the atmosphere prevailing at the times of observation. 

Fig.~\ref{3d-demo}a clearly shows a very localized increase in the
radial velocity\footnote{In the remainder of this article, velocity
  should be taken to mean the radial velocity, unless otherwise
  specified.} and the brightness. The typical extent of the areas
which show this phenomenon of strong localised brightness is
0.5\arcsec$\,$--$\,$1\arcsec\ in all the 193 regions identified in the
data. All regions are therefore well resolved, given a spatial
resolution of 0.18\arcsec. 

Data for each region have been characterised in terms of the following
properties.\\ 
\emph{(i)} the maximum velocity relative to the surrounding gas, \\
\emph{(ii)} the maximum brightness,\\
\emph{(iii)} the displacement in the plane of the sky between the
maxima in velocity and brightness.\\
\\
\emph{(i)} The maximum velocity of the flow relative to the
surrounding gas is found by taking the maximum velocity recorded in
the region of interest and subtracting the average velocity of the
surrounding gas. The latter may be estimated with a typical
uncertainty of 3-4\kms\ (see sect.~\ref{sec:datared}). The resulting flow
velocity contains this uncertainty. Thus these
velocities are less accurately determined than relative velocities
between adjacent bright regions. Using Fig.~\ref{3d-demo}a as an
example, the velocity of the surrounding gas is 8$\pm$3\kms. The tip
of the velocity peak is at $-$24$\pm$1\kms\ and thus the flow velocity
relatively to the surroundings is $-$32$\pm$4\kms\ (3$\sigma$). 
%

%

In 64 out of the 193 regions there is an increase in brightness similar
to that found, for example, in Figs.~\ref{3d-demo}a and b,
but with no corresponding velocity structure. An example is shown in
Fig.~\ref{3d-demo}c. These features will be referred to as features
with weak radial velocity components (WRVs). Fig.~\ref{3d-demo}c, which
is a typical such object, demonstrates that velocity variations are of the
order of less than 5\kms, with no clear velocity structure. This is
therefore used as a definition of WRVs, which are defined as those
objects with a localized maximum in brightness but which show no
velocity excursions from the surrounding gas greater than 5\kms. WRVs
according to their brightness and extent may arise from flows close to
the plane of the sky. However, some of  these features could also be
bright photodissociation regions (see sect.~\ref{sec:shockPDR}). In
this connection data for the great majority of WRVs are of
sufficiently high signal-to-noise that variations in velocity are real
and do not simply arise from uncertainties in velocity
determination.\\
\\
\emph{(ii)} The maximum brightness is the highest recorded
brightness within the confines of any region after the 3\,$\times$\,3 boxcar
smoothing mentioned in sect.~\ref{sec:datared}.\\
\\
\emph{(iii)} Spatial displacements of brightness and velocity
features are listed in Tables~\ref{table_p11}~--~\ref{table_cz}. There
exists a strong spatial correlation between the local peaks in
brightness and velocity: the distance in the plane of the sky between
the two peaks is typically less than the extent of either peak. That is,
they overly each other in the great majority of cases, with the
maximum in velocity lying within the area where the emission caused by
the flow is more than than twice the background emission.
Figs.~\ref{3d-demo}a and b illustrate this general
property that features of velocity and brightness are closely
associated in the plane of the sky. The use of these displacement data
to determine the orientation of gas motion in the plane of the sky is
discussed in sect.~\ref{sec:orient}.
\subsection{The Origin of the H$_2$ emission} \label{sec:shockPDR}
The mechanism by which the H$_2$ emission is formed is briefly
discussed here. This has already been considered
in \citet{v2001}, \citet{k2003} and Paper I. \citet{v2001} described
in detail, using the shock models of \citet{wilgenbus00}, how the
brightest regions in the Peak 2 could arise from magnetic (C-type)
shock compression of dense gas at pre-shock densities around
10$^6$\cm\ by shocks travelling at 30\kms. \refe{Using shock models
  and limiting their investigations to Peak 2, \citet{v2001} sought to
  find if quantitative evidence could be established that shocks have
  collected enough gas and compressed it sufficiently to overcome the
  Jeans criteria. This work was therefore directed towards discovering
  if shock-induced star formation is an active mechanism. Later 
VLT observations have in fact shown the presence of a 140~AU disk in
L-band around the position of brightest emission (15.8\arcsec~E
-1.6\arcsec~N) where \citet{v2001} concentrated their attention. Therefore a
protostar would appear already to have formed at this position and the
emission observed was most likely due to outflow from a YSO rather
than the much earlier stage of shock induced compression envisaged in
\citet{v2001}. While other candidates in Peak 2 may be less compelling
in terms of emission brightness and the amount of material within a
restricted volume, the mechanism of shock-induced star formation 
could nevertheless be active elsewhere Peak 1 and Peak 2. 

The shock-induced mechanism leads of course to star formation at some
time 10$^{4-5}$ years in the future, assuming collapse on a free-fall 
timescale. In this connection, according to the
latest shock models \citep{bourlot02,flower03,k2003,k2006}, the brightest
emission ($>$10$^{-5}$\wmsr, $>$300 counts per 400 seconds) may readily be
generated by magnetic shocks of typically 25\kms\ impinging on dense
gas ($>$10$^6$~cm$^{-3}$), much as in \citet{v2001}. Shocks leading to
the formation of the fingers occurred recently ($\sim 10^3$ years ago, 
\citet{doi02}) and the associated shock in dense gas in inner regions
of OMC1, such as Peaks 1 and 2 may lead to further star formation,
again on the 10$^{4-5}$ year timescale.}

Both \citet{k2003} and Paper I addressed the question of the relative
importance of shocks and photodissociation regions (PDRs) in yielding
the observed H$_2$ emission brightness. \thetc\ generates a UV field,
at around 100~nm, more than 10$^5$ times greater than the standard
interstellar field. Detailed models for H$_2$ emission have been
devised, for example in \citet{storzer99}, for UV fields of this
intensity falling upon dense gas, including advection and
photoevaporation \citep{henney99}. These models show that the maximum
contribution to emission brightness in the H$_2$ \mbox{v=1-0 S(1)}
line from PDRs does not exceed  a few times 10$^{-6}$\wmsr, given a
line-of-sight normal to the PDR.  For this geometry, the PDR
contribution to the brightness is therefore no more than 10\% to 15\%
of the maximum observed here. The conclusion is that the brightness
due to photoexcitation is a generally minor, though non-negligible
contributor to the H$_2$ emission observed here. This issue is
discussed in more detail, with reference to specific regions, in
\citet{k2003}.

\refe{Possible caveats to the above could
be provided by young highly obscured O-stars around which the UV field would be
even higher than the standard Orion value and the H$_2$ would be of
very high density. However the resulting H$_2$ 
emission would also be highly obscured from our view. Low-mass YSOs of
class 2 or 3 may also contribute with 100~nm emission, but a
comparison with the catalogues of stellar sources in \citet{muench02}
and \citet{lada04} (see sect.~\ref{sec:sfrs} and fig.~\ref{starplot})
shows that stellar objects of class 0 or I are only 
found near strong H$_2$\vone emission in a few cases}. Note that with a few
exceptions all of the areas examined in this work have a maximum
brightness which is 25\% or more of the observed maximum brightness,
and all regions showing clear evidence of flows show brightness above
the 15\% limit mentioned above for \refe{PDR contributions. Thus PDR
  mechanisms are unable to  reproduce the observed brightness of H$_2$
  emission.} This issue is re-examined briefly in the discussion of
weak radial velocity components (WRVs) in sect.~\ref{sec:d-edge} where
other geometries are considered in which PDR activity may play a more
significant role.

\section{Results} \label{sec:results}
\begin{sidewaysfigure*}
\resizebox{0.95\hsize}{!}{\includegraphics{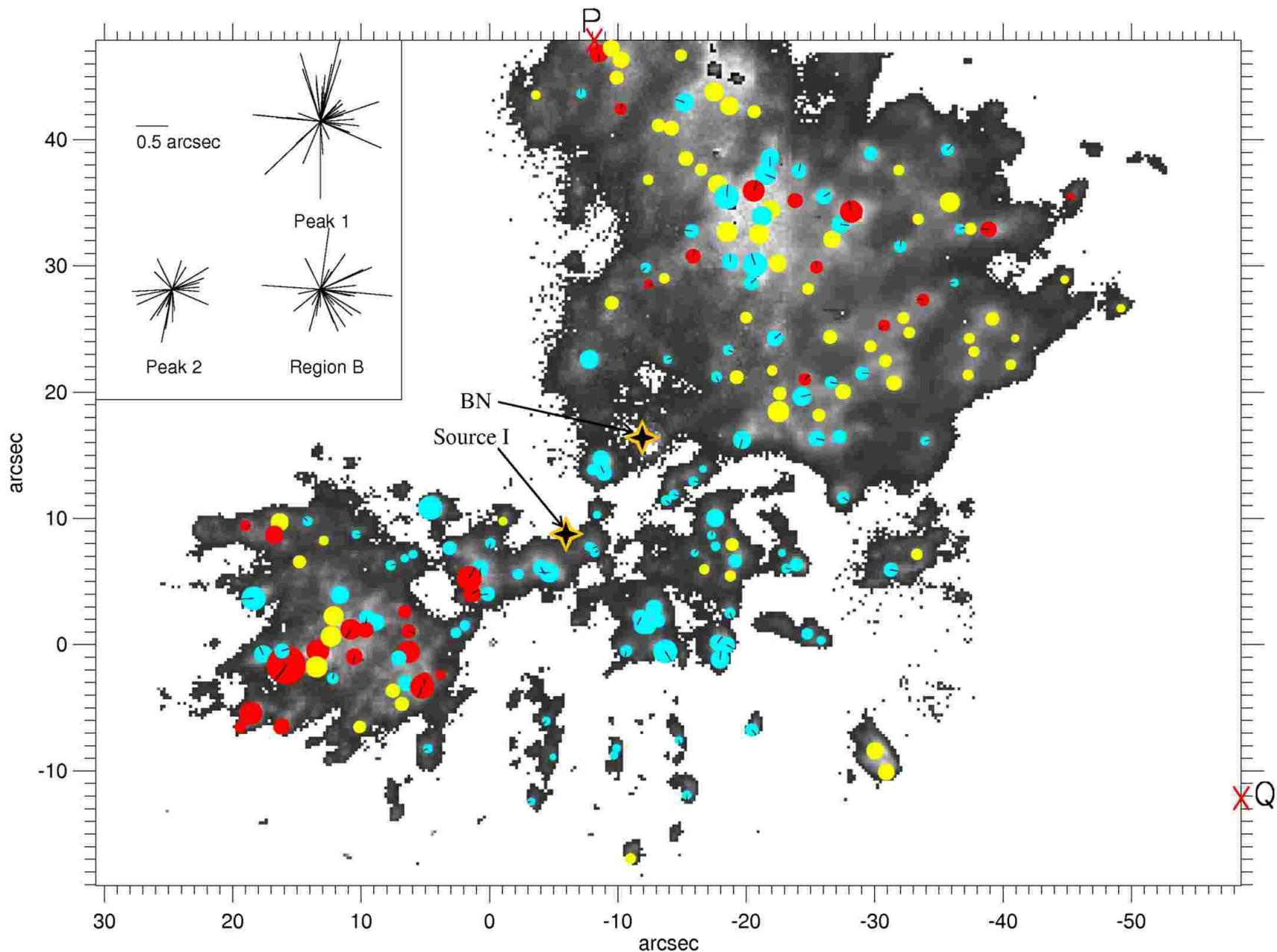}} 
\caption{H$_2$ v=1-0 S(1) emission at 2.121~$\mu$m (gray
  scale). Circles show all 193 features identified in the present work
  in this line. The centre of each circle marks the position of
  maximum brightness of a feature, and the radius of the circle is
  proportional to the maximum brightness. The colour of a circle shows
  whether a flow is red-shifted (red circles), blue-shifted (blue
  circles) or with no clear velocity feature, WRVs -- see text,
  (yellow circles). The position angle of the flow, if determined, see
  sect.~\ref{sec:orient}, is shown by a black line in the circle. The
  star diagrams in the upper left hand corner show all position angles
  in a given region. The length of the lines in each star diagram
  corresponds to the displacement between maximum brightness and
  maximum velocity. Points P and Q on the main figure axes define the
  orientations of the line of WRVs across Peak 1 (see
  sect.~\ref{sec:f-edge} and sect.~\ref{sec:d-edge}).}
\label{intplot}
\end{sidewaysfigure*}

To illustrate the nature of the data collated in
Tables~\ref{table_p11}~--~\ref{table_cz}, all features have been
plotted in Fig.~\ref{intplot}, where each circle marks an individual
feature. The radius of the circle is proportional to the maximum 
brightness, and the colour shows whether features are red-shifted (red
circles), blue-shifted (blue circles) or if WRVs were
found to be present (yellow circles). 
%
%
\subsection{Flow velocities} \label{sec:f-stat}
%
\begin{figure*}
\begin{center}
\resizebox{0.95\hsize}{!}{\includegraphics{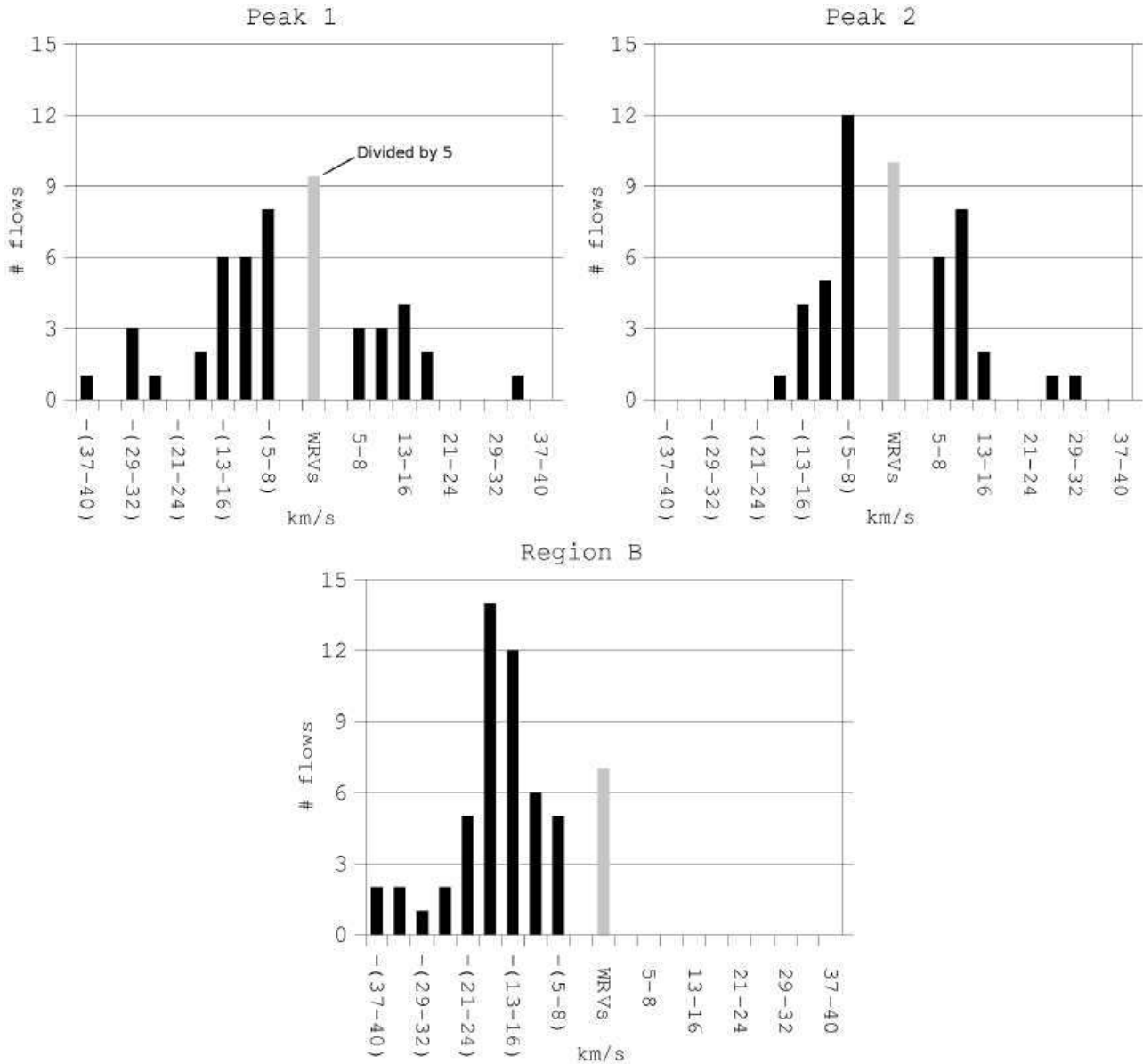}}
\caption{Histograms showing the number of flows with a given radial
  velocity relative to the velocity of the immediate surroundings in 
  each of the three main regions. Velocities are binned over 4\kms,
  except for WRVs which are inherently binned over 10\kms (see
  sect.~\ref{sec:findflows}) Weak radial velocity features (WRVs) are
  shown as gray-shaded bars.}
\label{stat-vel}
\end{center}
\end{figure*}
\begin{figure*}
\begin{center}
\resizebox{\hsize}{!}{\includegraphics[angle=90]{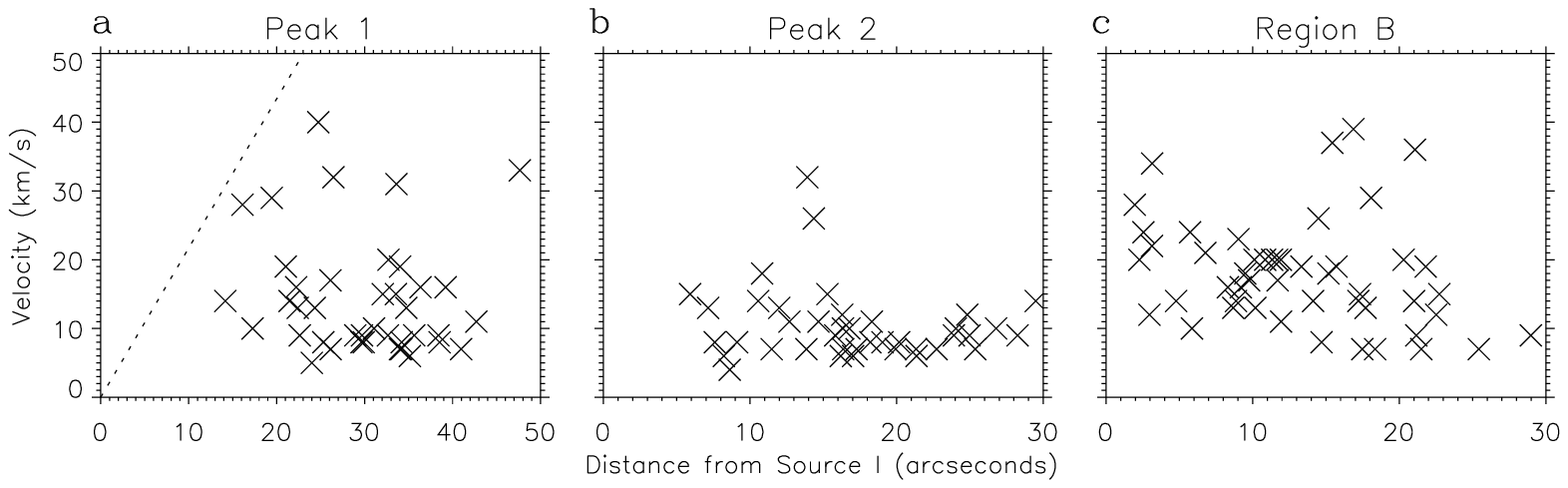}} 
\caption{\refe{Hubble diagrams showing velocity versus distance from
    Source I for each of the 3 regions. In \textbf{a}, the dotted line
    corresponds to an outflow of age 1000 years corresponding to the
    age of the OMC1 fingers according to \citet{doi02}. Note that the
    WRVs are not included in these diagrams since these have no
    measureable radial velocities.}}
\label{hubble}
\end{center}
\end{figure*}

%
Figure \ref{stat-vel} shows the number of flows for a given velocity
for Peaks~1 and~2 and Region~B. Several properties of the flows are
apparent:\\ 
\\
\emph{(i)} The fraction of red-shifted flows is similar in Peaks 1
and 2, with 33\% in Peak 1 and 45\% in Peak 2. There are thus overall
fewer red-shifted than blue-shifted flows. The detection of bright
H$_2$ emission is strongly affected by the local dust optical
depth. In dense regions associated with star formation the near-IR
may be strongly obscured \citep{rosenthal00}. Since the densities may
exceed 10$^7$\cm, one hundred to a few hundred AU depth of
material is sufficient to obscure H$_2$ emission. Thus material is
preferentially observed emerging from the obscuring dust, rather than
retreating into it, that is, blue shifted flows are expected to be
detected more readily than red shifted, as found. \\
\\
\emph{(ii)} Compared to Peak~2, Peak~1 has a large over-representation
of WRVs, as evidenced by Fig.~\ref{stat-vel}a and as is clear from
Fig.~\ref{intplot}. WRVs in Peak 1 \refe{may} have a different origin
than other 
features in Peaks 1 and 2, arising either from a concerted outflow
associated with the BN-IRc2 region and/or with a possible PDR
contribution. This is discussed in sect.~\ref{sec:f-edge} and
sect.~\ref{sec:d-edge}.\\ 
\\
\emph{(iii)} the total number of flows in Peak 1 and Peak 2 is 40 in
each case, excluding WRVs since these may represent in general a
different class of object. Since Peak~2 is less than half the size 
of Peak~1, the density of flows is correspondingly larger in Peak~2.\\
\\ 
\emph{(iv)} In Region B all flows are blue-shifted and the velocities
are distributed around an average value of $\sim-18$\kms\ with a
spread of 8\kms.\\
\\
\emph{(v)} For both Peak 1 and Peak 2 the red- and blue-shifted flows appear
clumped around $+10$\kms\ and $-10$\kms. \refe{Recall that velocities
  are relative to the v$_{lsr}$. Thus this symmetric distribution is
  the result we would expect if the origin of the flows is internal to
  OMC1, be it turbulent motion or outflows from protostars. The
  velocity dispersion of protostars is only a few \kms\ in the Orion
  Nebula \citep{odell01,vanaltena88}. Hence if H$_2$ emission arises
  to some extent from outflows from protostars, one would expect to
  observe a distribution of velocities which reflects the range of
  outflow velocities, since these greatly exceed relative protostar
  velocities. Thus data for peaks 1 and 2 in fig.~\ref{stat-vel} are
  consistent with the presence of protostellar outflows.}
\\
\\
\emph{(vi)} Flow velocities lie between a few\kms\ and an upper limit
of $\sim$35-40\kms\ for the radial velocity.  Again, on the basis that some
of the flows may be associated with protostellar outflows, these
figures agree with results from \citet{davis01} where flow velocities
of 5 -- 40\kms\ are recorded for a sample of 9 protostellar zones. The
upper limit is consistent with results in \citet{bourlot02} for the
critical velocity at which a C-type shock collapses to J-type, in gas
of pre-shock density 10$^5$--10$^6$\cm.\\
\\ \refe{
\emph{(vii)} Fig.~\ref{hubble} shows Hubble diagrams for each of the 3
regions, Peak 1, Peak 2 and Region B, plotted as the radial
velocities versus distance from source I. The dotted line in
fig.~\ref{hubble}a represents an age of 1000 years corresponding to
the age of the finger system as found by \citet{doi02}. There is no
evidence of Hubble-type behaviour in any of the observed regions.
Some of the data in fig.~\ref{hubble}a may contain close-in
representatives of the finger system, the great majority of whose
members lie outside our field as noted earlier. It is clear however that the
data provide no evidence that the observed shocks are part of the
concerted outflow from a common explosive event with which the fingers
are associated. One proviso is that Fig.~\ref{hubble} show only radial
velocities. Thus the lack of a distance-velocity correlation does not
in itself exclude an origin of the observed emission in some common
explosive event. Thus we may observe random azimuthal velocities
associated with flow in the plan of the sky, observed here as random
radial velocities. In fact we show, by appealing to orientation angles
of flows in sect.~\ref{sec:ori-ang}, that there is some directed flow
in Peak 1. We should also recollect the present discussion excludes
WRVs. In sect.~\ref{sec:d-edge}, we find that there may be a common
origin of the WRVs in Peak~1 and the bullets, again representative of
a directed flow to the NW.}
%
%
\begin{figure*}
\begin{center}
\resizebox{0.95\hsize}{!}{\includegraphics{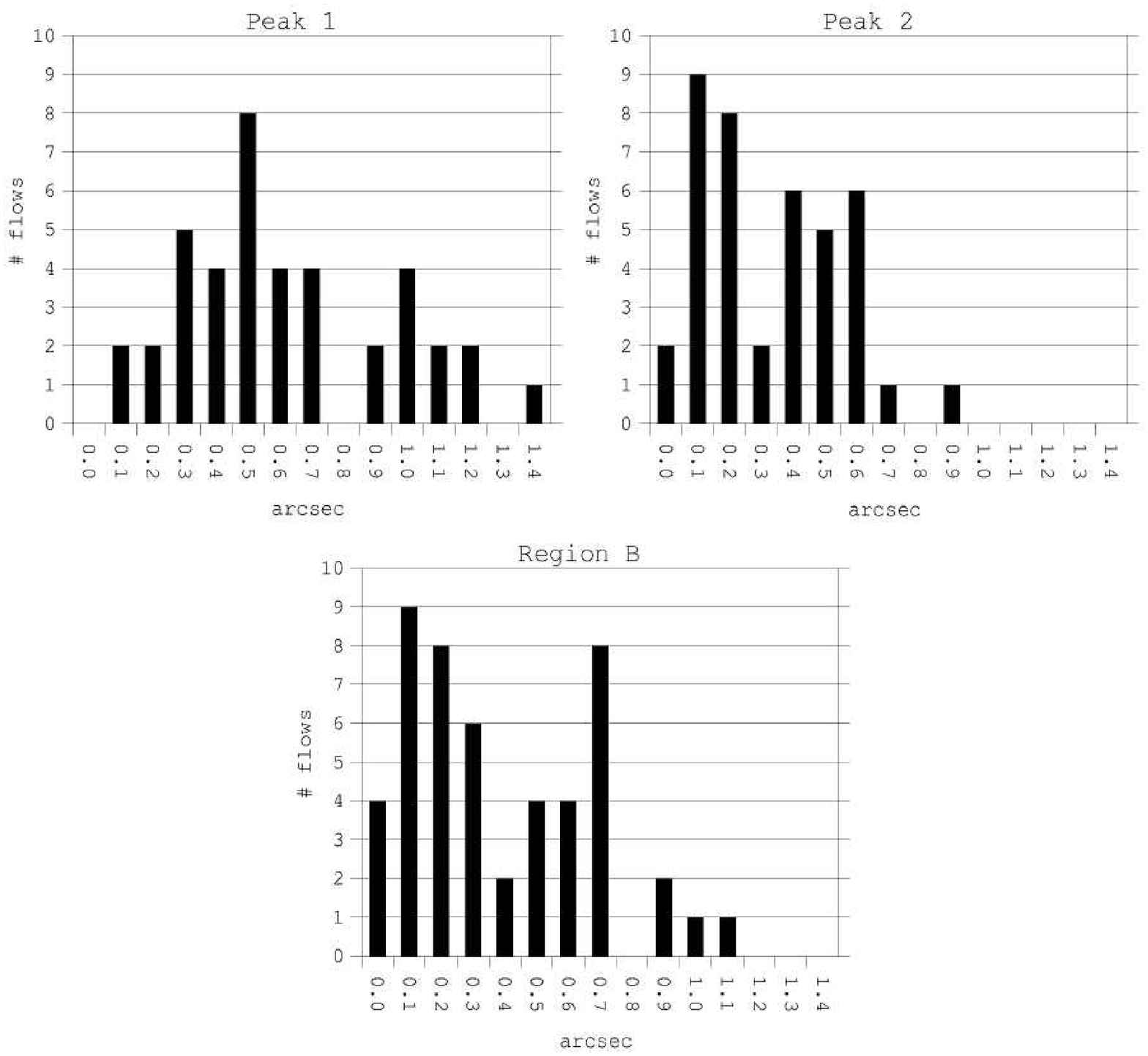}} 
\caption{Numbers of flows with a given displacement between the maxima
  in brightness and velocity for each of the three main regions.}
\label{stat-disp}
\end{center}
\end{figure*}
%

\subsection{Displacement of velocity and H$_2$ emission brightness}
  \label{sec:d-stat}
Fig.~\ref{stat-disp} shows the distribution of displacements between
associated maxima in brightness and in velocity for each region. We
find that for 2/3 of the flows the displacement between maximum
velocity and maximum brightness is $\leq$0.6\arcsec, and less than
10\% show a displacement of $\geq$1.0\arcsec. This should be compared
to the typical extent of a peak in velocity or brightness which is of
the order of 0.5\arcsec$\,$--$\,$1\arcsec. Data in
Fig.~\ref{stat-disp} quantify the earlier statement that a strong
correlation exists between the positions of maxima in brightness and
velocity (sect.~\ref{sec:findflows}), implying that these are connected
phenomena.\\
\subsection{Spatial distribution of flows with weak radial velocity
  components (WRVs)} \label{sec:f-edge} 
Several features are apparent in the spatial distribution of WRVs.\\
\\
\emph{(i)} WRVs comprise only a small proportion of the total number
of flows in Peak 2 while in Peak 1 they outnumber all of the flows
identified.\\
\\
\emph{(ii)} If the assumption is made that the WRVs arise from shock
excitation involving clumps of gas moving in or very close to the
plane of the sky, then there is an inhomogeneous distribution of
motion within Peak 1. This may be shown as follows. 

If the general flow pattern in Peak 1 were homogeneous in three
dimensions, then an estimate may be made of the number of flows that
are expected to fit into the category of WRVs.
%
\begin{figure}
\resizebox{\hsize}{!}{\includegraphics{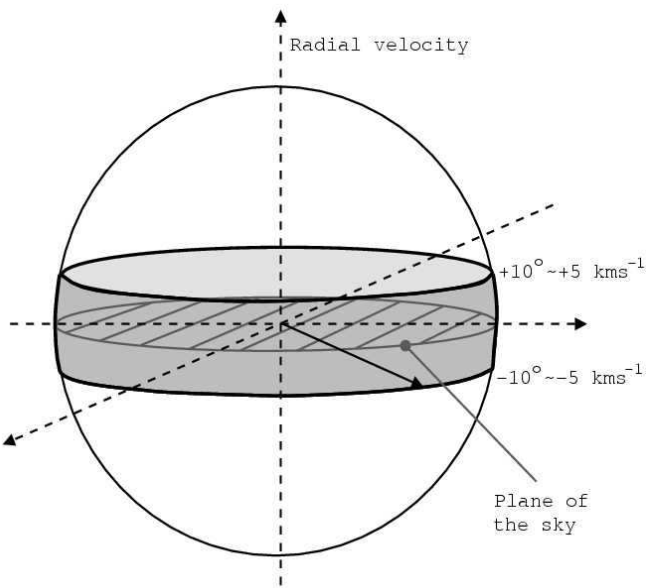}} 
\caption{Illustrating the ratio of the number of flows showing less
  than 5\kms\ radial velocity (WRVs) to the total number of flows: see 
  sect.~\ref{sec:f-edge} and Eq. 1.} 
\label{sphere}
\end{figure}
%
Given that typical shock speeds in OMC1 are $\sim$30\kms\ \citep[as
 suggested by results in][]{v2001}, then the limit imposed on the
radial velocity to define a flow as a WRV, v$_{radial} <$ 5\kms,
sect.~\ref{sec:findflows}, corresponds to flows moving within
approximately $\pm$10\degr\ of the plane of the sky. This is
illustrated in Fig.~\ref{sphere}.
For a homogeneous distribution of flow direction, the number of flows
designated as WRVs relative to the total number of flows should
equal the relative area of a band on a sphere around the equator from
$-$10\degr\ to $+$10\degr\ ($=\pi$/18 rad.): 
\begin{equation}
\frac{^{\textrm{no. of flows in the}}_{\textrm{plane of the
      sky}}}{\textrm{total no. of flows}} =
      \frac{\int_{-\pi/18}^{\pi/18} 2\pi \cos \theta d\theta}{4\pi}
      \approx 0.17 
\end{equation}
In Peak~2 a total of 50 flows (including WRVs) are found and thus in a
model of homogeneously directed flows, 8--9 WRVs are expected. This
agrees well with the 10 examples observed. The same model 
applied to Peak~1 suggests that $\sim$15 WRVs should be seen. However 47
WRVs are observed in Peak~1. The above results do not change
significantly if a typical shock speed of 25 or 20\kms\ is assumed
instead of 30\kms. Thus the flows do not form a spatially homogeneous
sample in Peak 1 but lie preferentially in the plane of the sky and
appear disproportionately as WRVs.\\ 
\\
\emph{(iii)} The data in Fig.~\ref{intplot} show not only that there is a
considerable excess of WRVs in Peak~1 but also that there is the
appearance of a concentration of WRVs along a direction on the
eastern side of Peak~1 from \mbox{$-$8\arcsec~E, 47\arcsec~N} to
\mbox{$-$31\arcsec~E, 21\arcsec~N} along the line connecting points P
and Q in Fig.~\ref{intplot}. This line (or slit) is $\sim$30\arcsec\ long but
only a few arcseconds wide. To assess whether this is a genuine
feature of the data or merely a trick of the eye, a slit of length
30\arcsec\ and width 2.5\arcsec\ has been placed at all positions and
for each position at all angles over the data for Peak~1 in
Fig.~\ref{intplot}. In each case the number of WRVs has been counted
within the slit. Figure~\ref{findedge} shows the number of WRVs in the 
slit at the optimum angle in the sky to pick up the maximum
number of WRVs, that is, 140\degr\ measured clockwise from N, as a
function of the perpendicular distance of the slit from TCC0016. This
corresponds to the slit being placed along lines parallel to the line
connecting points P and Q in Fig.~\ref{intplot}. The presence of a
very clear maximum maximorum indicates that there is a genuine
concentration of WRVs at the position designated. There is thus some
kind of ''front'' of emission features at this position.\\
\\
\emph{(iv)} A further property distinguishes the WRVs from other features
in the field within this concentration or front just identified. An
investigation of the morphology of the H$_2$ emission 
in these features reveals that $\sim$2/3 of these objects show a very
much more broken up morphology than that typically seen in features
elsewhere in Peak 1 and Peak 2, for which emission is clearly
associated with radial motion. Moreover, even in other WRVs, outside
of this concentration, such a broken up morphology is only seen in
10-15\% of the cases. The broken-up morphology associated with the
concentration of WRVs is illustrated in Fig.~\ref{ontheedge}b. This may be
compared to the relatively uniform increase towards a single maximum
seen in most other WRV objects such as the WRV shown in
Fig.~\ref{ontheedge}a. The origin of broken-up structure is briefly
considered in sect.~\ref{sec:d-edge}.
%
\begin{figure}
\resizebox{\hsize}{!}{\includegraphics{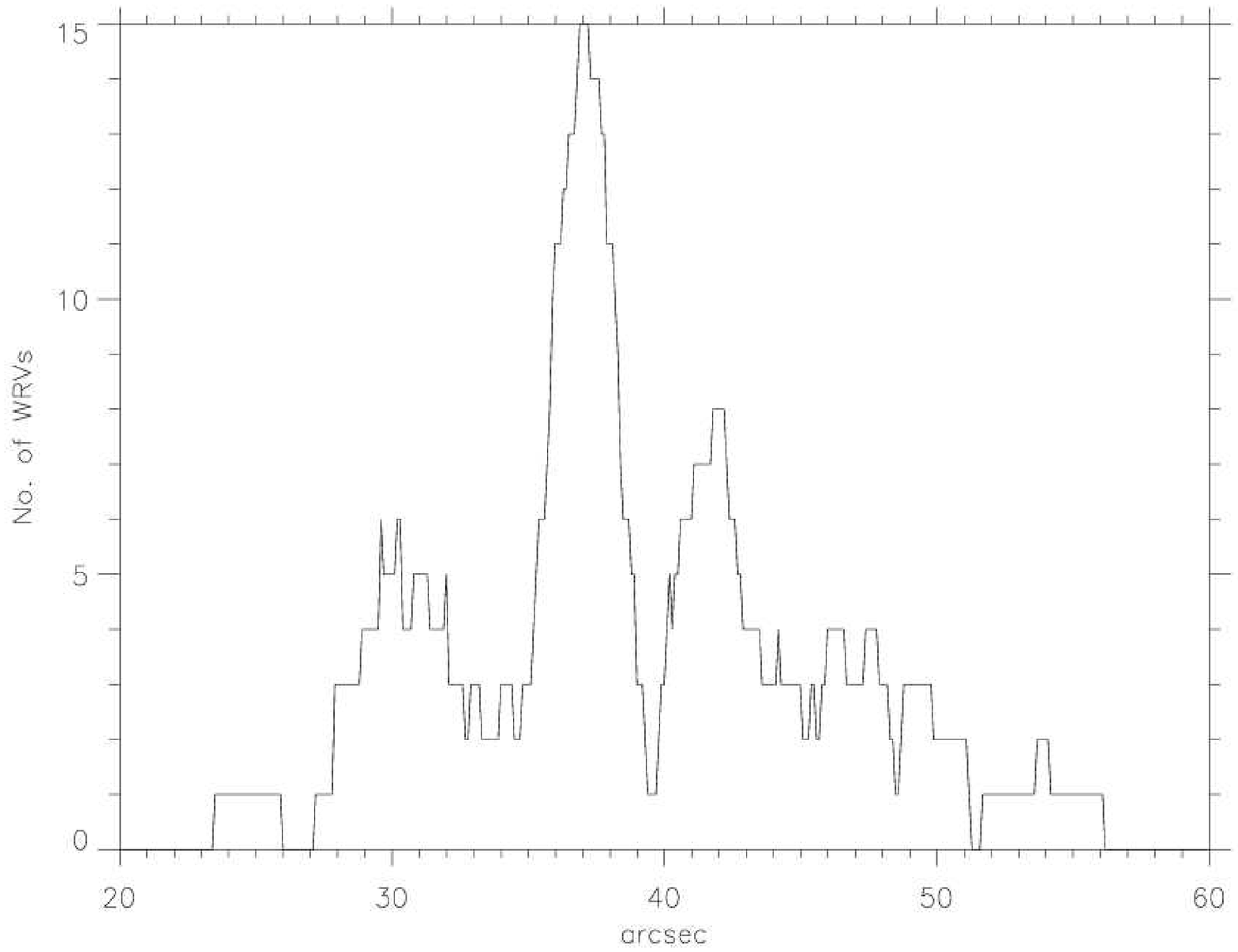}} 
\caption{The number of WRVs in a slit placed across Peak~1 at an angle
  of 140\degr\ measured clockwise from N as a function of perpendicular
  distance between the slit and TCC0016. The position of the peak at
  37\arcsec\ corresponds to a slit placed along the line connecting
  points P and Q in Fig.~\ref{intplot}.}
\label{findedge}
\end{figure}
%
\begin{figure}
\resizebox{\hsize}{!}{\includegraphics{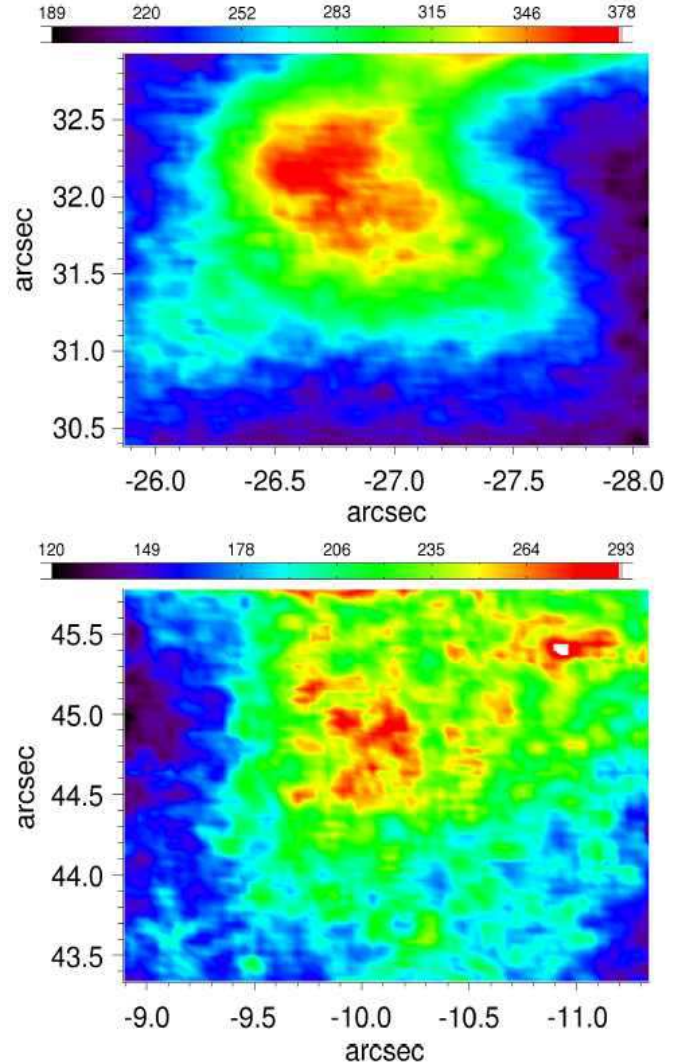}} 
\caption{\textbf{a} Flow no. 1-57: see Table~\ref{table_p11}. A
  smoothly structured WRV situated in Peak 1 but not in the flow front
  described in the text: sect.~\ref{sec:f-edge}. \textbf{b} Flow 
  no. 1-7: see Table~\ref{table_p11}. A highly structured WRV feature
  lying in the flow front. \refe{Colours denote brightness as
  shown on the colour bar above, in counts per 400 sec.}}
\label{ontheedge}
\end{figure}
%

%

\subsection{Grouping of flows} \label{sec:f-grouping}
\begin{figure*}
\resizebox{\hsize}{!}{\includegraphics[angle=90]{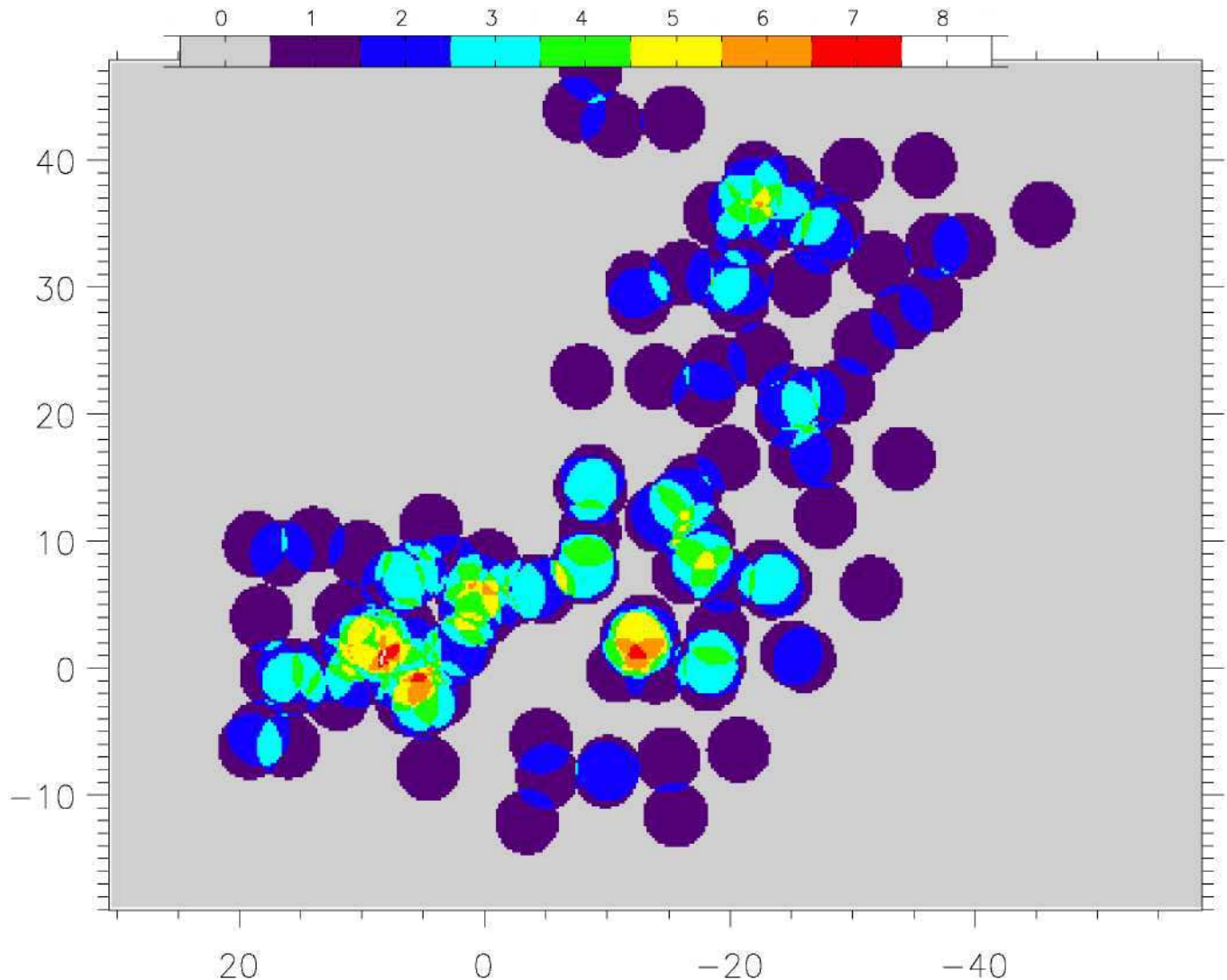}} 
\caption{Spatial density of emission features associated with flows,
  showing the number of flows within $\sim$1200~AU of any given
  point. The colour scale illustrates the number of overlapping
  features at any point. See sect.~\ref{sec:f-grouping} and
  sect.~\ref{sec:sfrs}. Axes are labelled in arcseconds.}
\label{density-blue-red} 
\end{figure*}

In the present section all flows in Fig.~\ref{intplot} are considered
save WRVs. Fig.~\ref{density-blue-red} shows how many blue- and
red-shifted flows lie within a distance of 1200~AU (2.6\arcsec) from 
any given point, where each flow is represented by a circle of radius
1200~AU. A radius of 1200 AU was chosen because a region of 2000-2500~AU
is a typical scale for a star forming core in the early stages of star
formation, e.g. \citet{bate03,delgado-donate04} and this is also the
scale where \citet{g2006a} find deviations in the turbulent velocity
structure, showing the presence of a preferred
scale. Fig.~\ref{density-blue-red} illustrates whether the flows have
a tendency to be grouped together or are spaced equally across the
entire area. Peak~1 and Peak~2 show different character. In Peak 1 the
flows are distributed rather evenly across the  entire area, with only
one  position showing a large number of flows in a small area, around
the position $-$22\arcsec~E, 36\arcsec~N. In Peak~2, flows are more
grouped with some areas showing 6-8 flows within an area of $\sim$2500
AU in diameter around positions 1\arcsec~E, 5\arcsec~N and 9\arcsec~E,
1\arcsec~N. Taken together with the
observation, reported in sect.~\ref{sec:f-stat}, that the spatial
density of flows is approximately twice as high in Peak 2 as in Peak
1, this points towards Peak~2 being a more active region compared to
Peak~1: see also sect.~\ref{sec:sfrs}. 
\subsection{Further characteristics of the outflow in Region~B}
  \label{sec:f-outflow} 
\begin{figure}
\resizebox{\hsize}{!}{\includegraphics{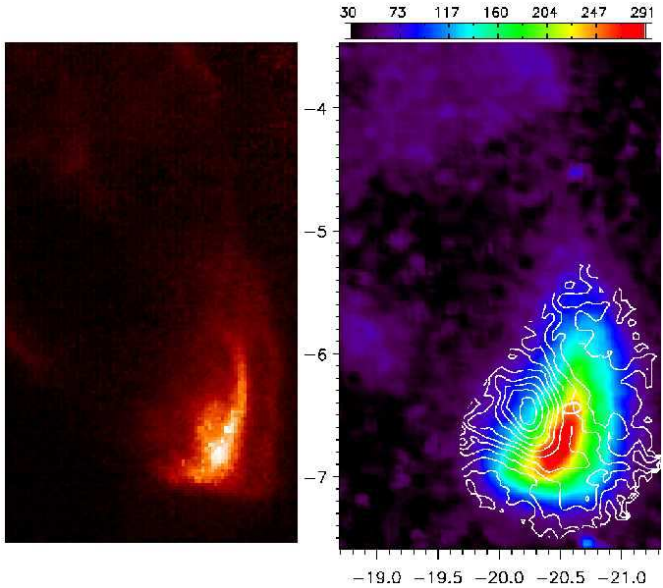}} 
\caption{Left hand panel: a VLT image of flow no. B-43,
  Table~\ref{table_cz}, taken with the NAOS-CONICA adaptive optics
  system, with a spatial resolution of $\sim$35~AU:
  see \citet{lacombe04}. Right hand panel: the same object obtained
  with the GriF instrument on CFHT, with a spatial resolution of
  70~AU. Contours show velocity and colours show brightness in counts
  per 400 sec. The maximum in velocity is -36\kms\ at~-20.2\arcsec~E,
  -6.5\arcsec~N. The interval between contour lines is 4\kms.}
\label{bowshock}
\end{figure}

As noted, all flows are blue-shifted in Region B and the mean of the
velocities is significantly greater than in Peaks~1 and 
2. Moreover Region~B contains, among the 49 objects in total, five
clear examples of bow shocks, based on the morphology of the
H$_2$ emission. An example is displayed in Fig.~\ref{bowshock}, the
left-hand panel of which shows a NAOS-CONICA adaptive optics image
obtained with the VLT \citep{lacombe04}, lacking velocity data but
showing the morphology in graphic detail, and the right hand side
showing the GriF results, including velocity data as contours.

\begin{figure}
\resizebox{\hsize}{!}{\includegraphics{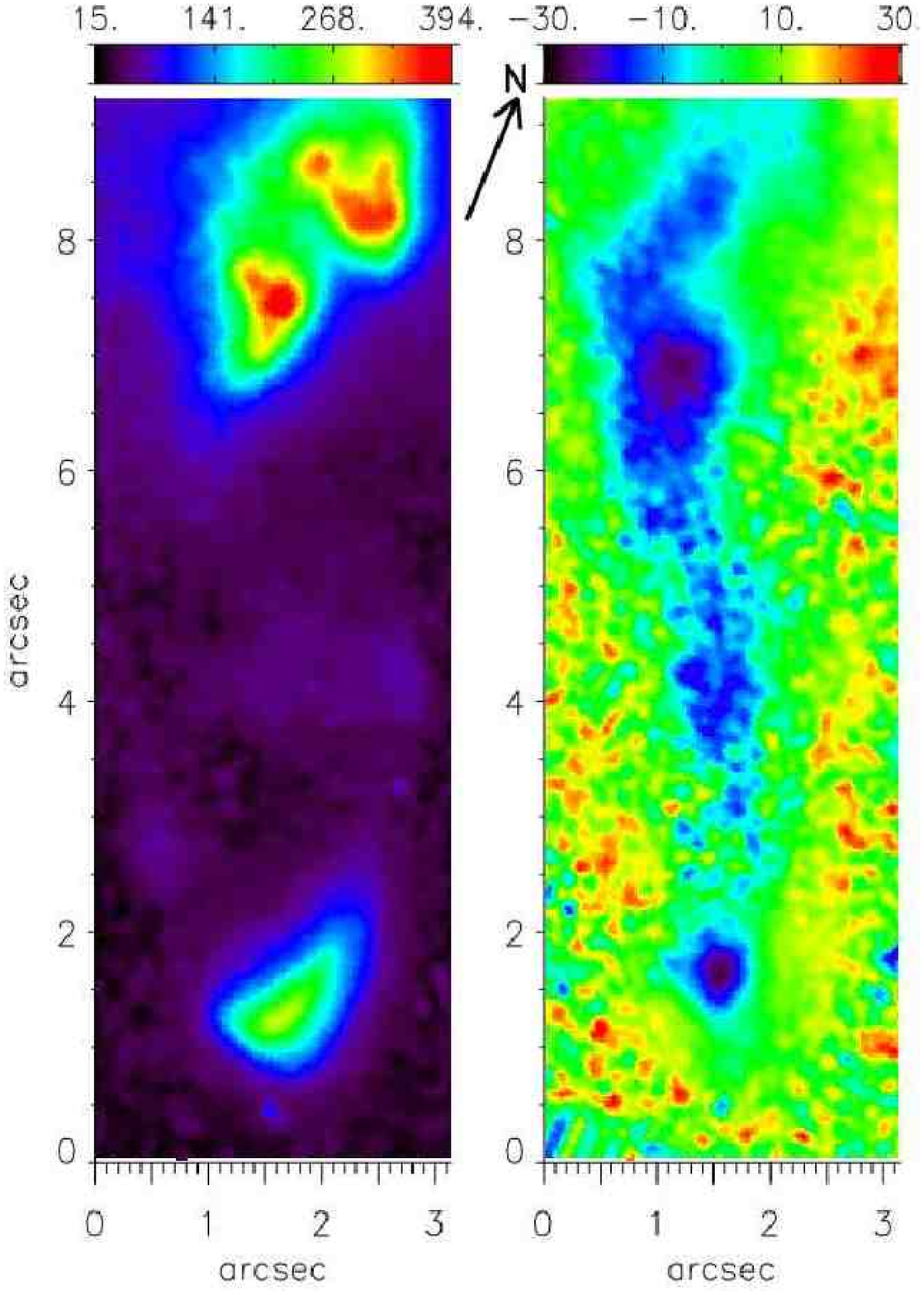}} 
\caption{GriF data for three bow shocks within the outflow
  Region~B. Left-hand image: H$_2$ emission brightness in v=1-0
  S(1). Colours denote brightness in counts per 400 sec. Right hand
  image: velocity data for the same field. The colour scale show
  velocity in\kms. The orientation 
  of the images is such that north points at an angle of 24\degr\
  measured clockwise as indicated on the figure. The objects involved,
  lying 15\arcsec\ to 20\arcsec\ SW of Source I (see
  Fig.~\ref{intplot}) are B-43 to the south (see also
  Fig.~\ref{bowshock}) and B-35--37 in the north, see
  Table~\ref{table_cz}.} 
\label{ridge}
\end{figure}

The structure in Region~B appears at first to consist of
disconnected fragments of emission. However on closer examination it
is found that there is widespread weak emission connecting some of the
prominent bright features shown in Fig.~\ref{intplot}. An example is
shown in Fig.~\ref{ridge}. A ridge of very weak emission, at the level
of 30-60 counts per 400 seconds, corresponding to velocity
uncertainties of 3 -- 6\kms, can be identified connecting three
bow-shocks showing blue shifted velocities between $-10$ and
$-$30\kms. Thus the outflow region represents a concerted bulk motion,
though diffuse in places.
%
%
%
%
%
%
\section{Discussion} \label{sec:discuss}
Gas flows within OMC1 are of two types. The first is large scale
flows. Two such flows may be identified in OMC1. There was a
large-scale outburst $\sim$1000 years ago \citep{doi02} roughly in
the plane of the sky, which has lead to the subsequent
formation of fast moving fingers and bullets in the outer reaches of
OMC1 to the NW e.g. \cite{lee00,doi02}. The WRVs concentrated in Peak 1
may be part of this scenario, as we discuss in
sect.~\ref{sec:d-edge}. The flow identified here in Region~B
constitutes the other large scale feature. By contrast, this is
largely out of the plane of the sky. The relationship between these
two outflows is not presently clear. The outflow in Region~B is
discussed in detail in sect.~\ref{sec:d-outflow}.

The second type of feature is small-scale flows, associated with local
low mass star formation. Flows in this case, taken as a body, would be
expected to show no overall preferential direction of motion within
the plane of the sky, with each site or group of sites acting
independently of any other. This point is examined in
sect.~\ref{sec:orient} below. 

\subsection{The physical nature of shocks in OMC1 and their
  orientation in the plane of the sky} \label{sec:orient} 
This section considers how data for the displacement of the maxima in
velocity and brightness in any chosen feature, recorded in the final
column of Tables~\ref{table_p11} -- \ref{table_cz}, may be used for
the determination of the orientation of shocks in the plane of the
sky. To enable this interpretation, it is necessary to consider the
brightness and velocity structure of C-type shocks and the spatial
relationship between the regions of maximum velocity and maximum
brightness in H$_2$ emission.

\subsubsection{A brief overview of shock structure}
\label{sec:shockstructure} 
\refe{The following gives a simplified description of the general
  events at play in a shock. This description is only applicable to
  features in OMC1 in broad outline but presents a physical picture
  which should prove helpfulin interpreting our observations.} We
envisage the shock forming as follows: in a time dependent picture,
starting at time zero, before the shock has been created, a jet of
  gas, from hereon for simplicity considered continuous, is ejected
  from a protostar and collides at a super-Alfvenic relative velocity 
  with a dense clump of gas in the surrounding medium. The jet slows,
  energy is given up to the system as heat, and H$_2$ in the impacted
  gas is sufficiently excited that emission takes place in IR vibrational
bands. C-type shock simulations show that the gas that emits most
strongly is moving at a velocity of about ten\kms\ less than that of
the initial jet. For example, there may be a jet with an initial
velocity of 20\kms\ impinging on gas with a pre-shock density of
10$^6$\cm. In the steady state which eventually forms, the heated gas
emits most brightly in H$_2$ when the flow has slowed to $\sim$10\kms,
according to C-type shock models \citep[ L.~Kristensen, private 
communication]{wilgenbus00,bourlot02,flower03}. The gas however continues 
to emit in vibrationally excited states of H$_2$ at even lower
velocities.

As the initial collisional process described above takes place, more
gas is entering the system from the continuous jet. This newly arrived
portion of the jet encounters gas which is moving several\kms\
more slowly. Where these two portions of the flow meet, there forms a
structure called the ''Mach disk'', which is itself a shock, given
that the relative flow speeds are super-Alfvenic. Thus there are, to
use the model of \citet{raga93}, effectively two shocks, one at
the Mach disk, and one around the apex of the system where the gas is
impacting the ambient material. The gas flows in front of and around
the outside of the Mach disk, forming a bow structure with so-called
''working surfaces'' at which weaker shocks form in 3D around the
region of the Mach disk. This is illustrated in Fig.~1 of
\citet{raga93}. In all parts of this flow, the gas cools and emits:
hence the structure is seen in H$_2$ emission. 

This general model finds support in our observations through the fact that
the physical apex of the shock, where it can be identified, is not
where the highest velocity is found. The physical tip of the shock has
slowed down, following momentum conservation taking account of the
compression of the gas ($\rho_1$v$_1$ = $\rho$$_2$v$_2$). From our
observations we find that typically the tip of a bow shock is at a velocity 
between 1/2 and 1/3 or lower of the maximum velocity. When a magnetic
field is present, as here, the working surface becomes a greatly
extended zone whose dimensions may be several hundred AU or
more, but are typically 50-60~AU in high density regions
of (say) 10$^6$\cm\ \citep{wilgenbus00,bourlot02,lacombe04}. 

Following the above model, the Mach disk is in general the point of
highest velocity in the shock relative to the surrounding gas,
assuming a constant jet playing on the 
medium. The question arises, is the Mach disk brighter or less bright
in H$_2$ emission than the portion of the gas in front of it, which is
moving more slowly? One might suppose that the brightest emission
would tend to lie in front of the Mach disk, since it is in front of
the disk that bulk energy is being turned into heat. However in C-type
shocks, as in OMC1 \citep{v2001,k2006}, the extended shocked zone may
well cause the Mach disk to merge with the bow region. A quantitative
understanding of the velocity-brightness structure of C-type shocks
requires high  density models which include both chemistry and at
least 2D hydrodynamics. At all events no double maxima in the
brightness structure of shocks are seen in the present observations.

\refe{In the above we assume a stable continuous system generated
  through a stable jet of constant speed. The exciting jets cannot be
  readily identified as distinct structures in OMC1 but where they are
  observed in class 0 and class I object, protostellar jets may appear
  time-variable in their velocity
  \citep{bontemps96,stanke00,bally00,larson03}. This would produce a
  more complex shock structure with several emission maxima, but does
  not in itself involve any shock formation mechanism fundamentally
  different from that outlined above. One point to be noted regarding
  variable jets speeds is that this can result in both forward and
  backward facing, or reverse, shocks. Reverse shocks are thought to
  be formed when an initially slow wind becomes a fast wind which then
  overtakes and compresses the material ahead of it. This gives rise
  to clumped material where the fast jet impinges on the slower
  jet. In the limit of a variable jet, the jet consists of spatially
  separated clumps of material each of which individually forms
  independent shock structures. This is illustrated in fig.~1 of
  \citet{hartigan89}.}

\refe{A further point is the relative dimension of the jet compared to
  that of any clump of material on which it impinges. In an idealized
  view, forward bows are formed when the jet radius is smaller than
  the dimension of the slow-moving medium and reverse bows form when
  dimension of the obstacle is smaller than that of the jet.}

\subsubsection{Determination of orientation angles} \label{sec:ori-ang}
On the basis of the previous section the brightest emission and the
highest velocity are expected to be closely spatially associated. The
shock travels in a direction from maximum velocity to maximum
brightness. This model will be referred to as the velocity to
brightness model. In some cases observations clearly support this
model. An example is shown in Fig.~\ref{bowshock}, object B-43,
Table~\ref{table_cz}. Here the maximum velocity is spatially displaced 
from the maximum brightness by 0.3\arcsec\ and the object gives every 
appearance of a bow shock travelling in the direction of maximum
velocity to maximum brightness. 
A counter-example to the above is however furnished by object 2-20 in
Table~\ref{table_p2}. Here the bow shape of the emission suggests
motion in a NW direction, whereas application of the 
velocity to brightness model for the shock motion indicates motion in
the SE direction. Furthermore, in areas with a high spatial density 
of flows, perhaps indicative of a multiple star forming region (see
sect.~\ref{sec:sfrs}), closely situated pairs of outflows where
one is red-shifted and the other blue-shifted are seen. However, for
all pairs of red- and blue-shifted shocks where position angles are
determined by the maximum velocity to maximum brightness flow
direction model (see below), the flows appear to be converging in the
plane of the sky. If any pair of flows formed a bipolar outflow, they
would clearly be expected to be found diverging, that is, moving apart
from one another. An example of this discrepant behaviour may be found
in Fig.~\ref{complexareas}a showing flows 1-76 and 1-81 (see
Table~\ref{table_p11}).

\begin{figure}
\resizebox{\hsize}{!}{\includegraphics{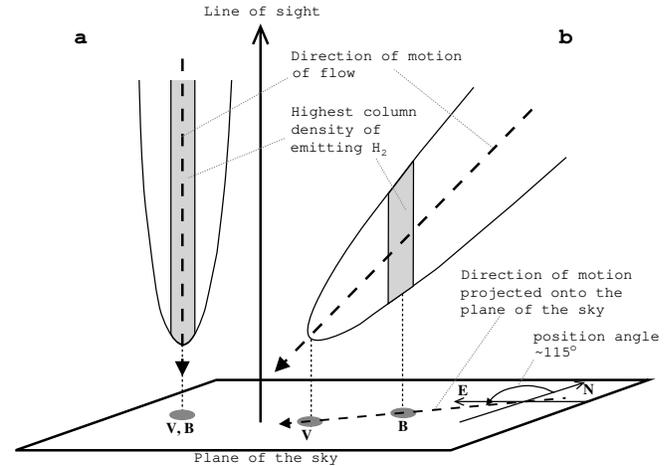}}
\caption{An illustration of the proposed geometry by which position
  angles are estimated in the column density model (see
  sect.~\ref{sec:ori-ang}). V's marks the position of the observed maximum
  in velocity and B's in brightness in the plane of the
  sky. \textbf{a} a flow moving in the line of sight: the maxima in
  brightness and velocity coincide. \textbf{b} A flow moving at an
  angle to the line of sight: the maximum in brightness will appear at
  some point behind the tip of the shock. A vector through the two
  maxima then gives the direction of the flow projected onto the plane
  of the sky, the position angle.}
\label{model}
\end{figure}

Failure of the velocity to brightness model will arise if
effects of optical depth are paramount in determining the
location of maximum brightness. The brightness of emission is a
function not only of the local efficiency of excitation associated
with local loss of kinetic energy, but also of the column density of
gas which is observed. Thus a shock travelling at some arbitrary angle
may appear brighter behind the position of maximum velocity by virtue
of its bow form, although locally the emission is weaker than at the
tip of the shock, where the latter is of low column density. This
model is illustrated in Fig.~\ref{model}, and was described in
Paper~I. This model will be referred to as the column density model.
%
%

\refe{The common feature of both models is that} the maximum in
emission and the maximum in velocity lie on a vector defining the
orientation of the shock in the plane of the sky. However, in the
absence of a full 2D or 3D C-type shock code for H$_2$ excitation
including chemistry, it is not possible to determine which processes
and thus which model may be relevant in any particular case, save
where the morphology clearly indicates a bow shock (as in
Fig.~\ref{bowshock}). Tables~\ref{table_p11} -- \ref{table_cz} quote
position angles measured east of north (i.e. counter-clockwise) based
on the velocity to brightness model where shocks are travelling in the
direction from the position of maximum velocity to the position of
maximum brightness in the plane of the sky. However, with a few
exceptions where the morphology of the flow would appear to dictate
the absolute position angle, e.g. fig.~\ref{bowshock}, these
position angles are uncertain by 180\degr\ with respect to absolute
direction as discussed above. The exceptions are marked with a star
and are found only in Table~\ref{table_cz} for the outflow region B. 
The uncertainty in the position angle is estimated to be
$\sim\pm$25\degr, decreasing with increasing displacement. The
uncertainty stems from the determination of the exact positions of the
maxima. A displacement of $<$0.25\arcsec\ of maximum brightness and
maximum velocity is close to the resolution 
limit, and small variations in displacement would have a large
influence on derived position angles. Position angles are therefore
only determined where the displacement is $\geq$0.25\arcsec.
%
\begin{figure*}
\begin{center}
\resizebox{0.9\hsize}{!}{\includegraphics{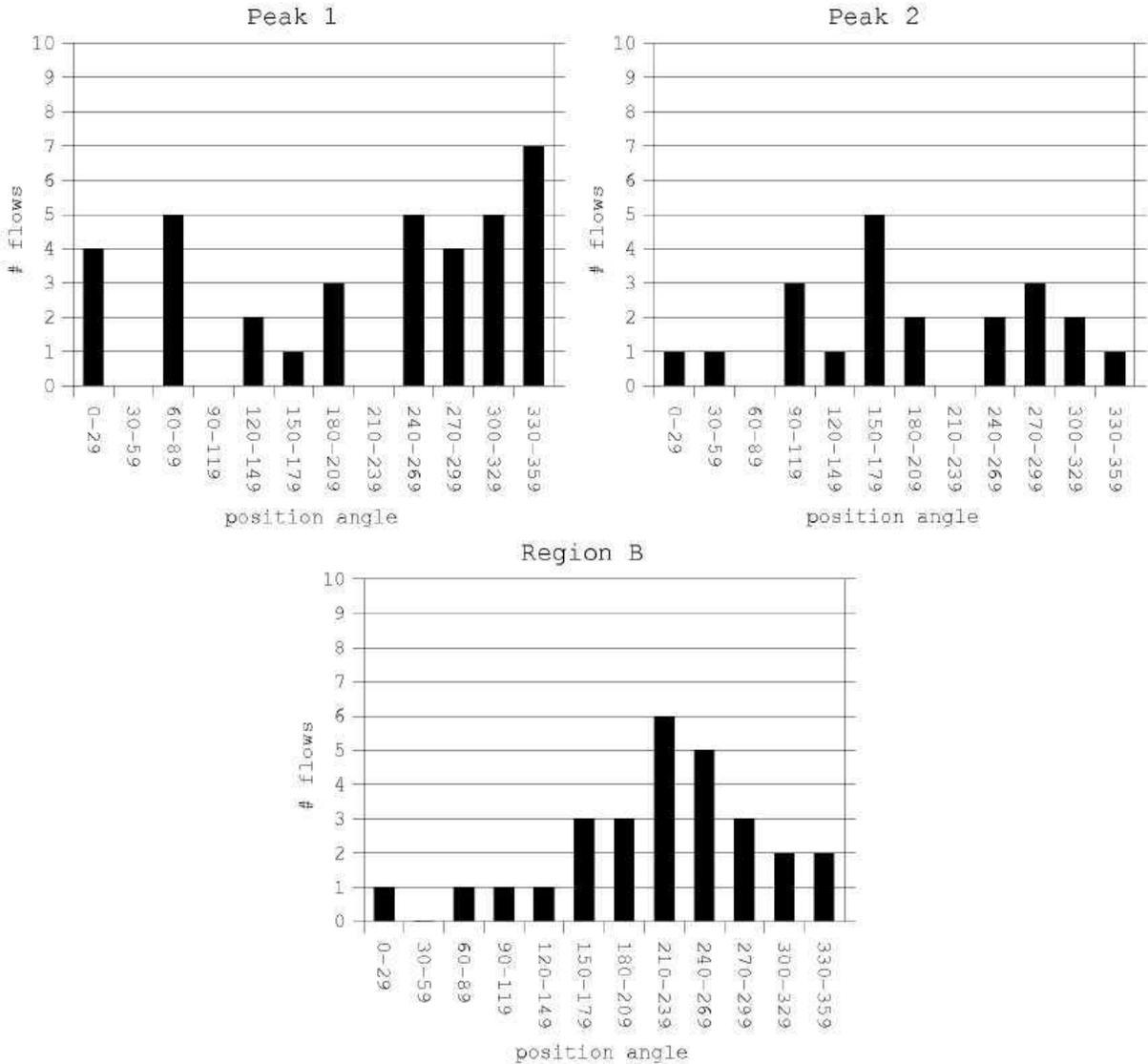}} 
\caption{Number of flows in each of the three main regions with
  position angles in a given range (see also star diagrams in
  Fig.~\ref{intplot}).}
\label{stat-pa} 
\end{center}
\end{figure*}
%

\begin{figure*}
\begin{center}
\resizebox{\hsize}{!}{\includegraphics{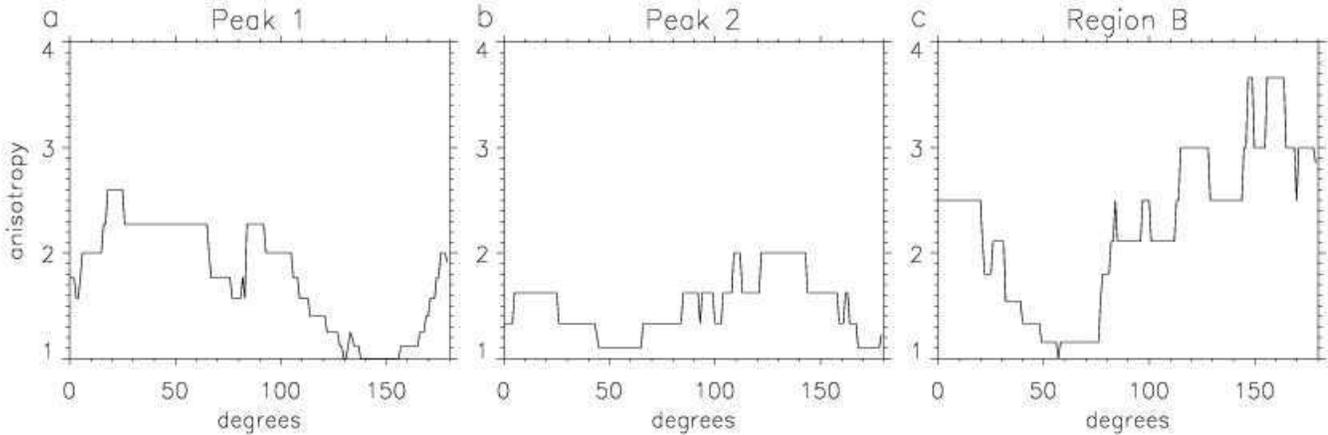}} 
\caption{The ratio between the number of flows on each side of a line
  through the star diagrams in Fig.~\ref{intplot} as a function of the
  angle between the line and north measured anti-clockwise from north. A
  high value suggests an anisotropy in the distribution of position
  angles across the line while a value of unity indicates an isotropic
  distribution.} 
\label{anisotropy}
\end{center}
\end{figure*}

An important consideration is whether the position angles are dominated
by a general outflow motion from the BN-IRc2 area or are approximately
isotropic in distribution in each main region (Peak~1, Peak~2 and 
Region~B). Fig.~\ref{stat-pa} show the distribution of position angles
binned over 30\degr\ intervals, with the reservation that individual
position angles may be incorrect by 180\degr\ (see above
discussion). In this connection, the assumption has been made that one
or the other model for position angles (velocity to brightness model
or column density model) will dominate in a given zone, given that a
single mechanism causing shocks predominates in any one zone, that is,
outflows from star forming regions or a large scale flow impinging on
the region (as in Region B). With this assumption, any preferred
directions can be 
found by looking for anisotropies in the distribution of position
angles. Rebinning the histograms in Fig.~\ref{stat-pa} to just two
bins of width 180\degr\ and then shifting the absolute positions of
the bins to cover all possible combinations will show anisotropies as
large differences in the number of flows in each bin. This corresponds
to placing a line across the star diagrams in the inset in
Fig.~\ref{intplot} and counting the number of flows on each side of
the line. The greatest difference between the numbers above and below
the line will appear when the line is perpendicular to the direction
of an overall motion, while there will be roughly equal
numbers when the line is parallel to any overall motion. If the line
is rotated (corresponding to the bins in Fig.~\ref{stat-pa} being
shifted) any preferred direction will appear as a maximum in the ratio
between the number of flows on each side of the line at an angle
perpendicular to the outflow direction. If there is no preferred
direction of motion the ratio should be approximately unity
independent of the angle of rotation. Fig.~\ref{anisotropy} shows this
ratio as a function of angle measured counter-clockwise from north for
each main region. The ratio is calculated so that it is always
$\geq$1.

Region B (Fig.~\ref{anisotropy}c) shows a
high degree of anisotropy around 140--160\degr\ and a minimum around
50--70\degr. As described above, this property with a maximum and a
minimum perpendicular to the maximum, indicates that the motions are
dominated by an overall outflow motion. In this case the motion is
along a line oriented $\sim$ 50--70\degr\ counter-clockwise from
north, that is, approximately NE-SW. The nature of this outflow in
Region B will be discussed further in sect.~\ref{sec:d-outflow}. 

Fig.~\ref{anisotropy}a shows a minimum
around 130--160\degr\ and a maximum at 20--50\degr. This is similar to
the structure observed for Region~B although the maximum is less
pronounced. This suggests that the observed flows in Peak~1 are 
also partly described by an outflow motion (towards the NW in this
case), but not to the same degree as in Region~B. Finally Peak~2
(Fig.~\ref{anisotropy}b) shows only very weak maximum-minimum
structure, with a distribution of position angles which is close to
isotropic. This suggests that the flows observed in Peak~2 are only
weakly affected by general outflow motion and tend rather to be
individual flows, such as might arise from local star formation.
%
%
\begin{table}
\begin{minipage}[t]{\columnwidth}
\caption{\refe{Good protostellar candidates, ''candidates'', (upper
    half of table) and weaker candidates, ''possible candidates'',
    (lower half) in the area of OMC1 covered in the present work,
    based on data taken from \citet{muench02}. For purposes of
    cross-reference, we have adopted the numbering of
    \citet{muench02}.}}
\label{table_psc}
\centering
\renewcommand{\footnoterule}{}  
\begin{tabular}{c c c c c}

\multicolumn{5}{c}{Protostellar candidates in OMC1} \\
\hline \hline
\multicolumn{5}{c}{Candidates}\\
\multicolumn{5}{c}{K$-$L $>$ 1.5 ;  J$-$H $>$ 1.0 or J$-$H undetermined.} \\
\hline
\citeauthor{muench02} & \multicolumn{2}{c}{Coordinates} &
\multicolumn{2}{c}{Colour indices} \\
numeration & \multicolumn{2}{c}{rel. to TCC0016} & K$-$L & J$-$H \\
\hline
00506 & -18.3 &-17.1 & 5.77 & ---  \\  
00543 &   9.5 & -7.2 & 3.29 & ---  \\  
00557 &  -0.9 & -5.0 & 2.32 &$>$1.82\footnote{\citeauthor{muench02}
  quote a detection limit of 18.15 in J band.}\\ 
00560 & -38.4 & -2.3 & 3.03 &$>$0.82$^a$ \\  
00569 & -24.0 & -0.1 & 3.57 & ---  \\  
00572A&  -3.6 &  0.6 & 1.52 &$>$2.67$^a$ \\  
00576 &  19.2 &  1.6 & 1.77 & 1.07 \\  
00578 & -14.6 &  1.6 & 4.42 & ---  \\  
00580 &   4.4 &  2.4 & 2.76 & 3.77 \\  
00583 &  -8.3 &  3.0 & 1.50 & 1.86 \\  
00596 &  14.6 &  5.9 & 1.53 & 1.11 \\  
00598 &  -8.4 &  6.3 & 2.99 & 3.30 \\  
00603 &  -0.6 &  7.5 & 2.22 & 1.02 \\  
00614A&  -2.9 &  9.4 & 2.81 &$>$4.65$^a$ \\  %
00645 & -17.6 & 17.1 & 1.82 & 1.32 \\  
00709A&  -9.2 & 34.7 & 1.99 & 1.25 \\  
00726 & -16.7 & 39.4 & 2.55 & 2.94 \\  
\hline 

\multicolumn{5}{c}{ }\\
\multicolumn{5}{c}{Possible candidates}\\
\multicolumn{5}{c}{No L band ; H$-$K $>$ 2.0 or K band photometry only.}\\
\hline
\citeauthor{muench02} & \multicolumn{2}{c}{Coordinates} & &  \\
numeration & \multicolumn{2}{c}{rel. to TCC0016} & H$-$K & M$_K$\\
\hline
00552 & -20.9 & -6.1 & 2.81 & 13.03 \\  
00554 &  -4.7 & -5.8 &$>$2.79\footnote{\citeauthor{muench02}
  quote a detection limit of 17.8 in H band.} & 15.01 \\  
00561 &  15.8 & -2.1 & 3.43 & 13.22 \\  
00570 &  -5.7 &  0.0 &$>$3.00$^b$ & 14.80 \\  
00585 & -10.7 &  3.4 & 3.19 & 10.96 \\  
00591 & -11.1 &  4.9 & 3.49 & 10.97 \\  
00594 &   1.5 &  5.2 &$>$3.85$^b$ & 13.95 \\  
00609 & -40.8 &  8.3 &$>$1.63$^b$ & 16.17 \\  
00610 &  -9.2 &  8.5 & 6.09 & 12.44 \\  
00634 &  -8.6 & 14.2 & 3.35 & 10.78 \\  
00665 & -17.1 & 21.7 & 2.04 & 11.01 \\  
00691 & -25.1 & 29.3 & 3.05 & 13.39 \\  
00695 & -37.4 & 30.4 & 3.13 & 15.59 \\  
00698 & -44.0 & 31.1 &$>$2.91$^b$ & 14.89 \\  
00713 & -47.7 & 36.1 &$>$2.99$^b$ & 14.81 \\  
\hline \hline
\end{tabular} 
\end{minipage}
\end{table}

\subsection{Protostars and Star forming regions in OMC1} \label{sec:sfrs}
Patterns of gas motion and spatial distribution of flows are examined
here for Peaks 1 and 2. Our discussion provides further evidence for
on-going low 
mass star formation in OMC1. The major source of energy injection into
OMC1 is at 0.1~pc (20000~AU) scales through outflows from massive
stars and as part of a large scale cascade of turbulent energy
\citep{g2006a}. Thus many of the flows, both large and small scale,
identified in the present work are an expression of the turbulence
inherent in the gas through massive stellar outflows from the BN-IRc2
complex \citep{shuping04,greenhill04b} and from source I \citep[see
  below;][]{greenhill04a}. However the gas dynamics may also be
influenced by the events which we focus upon here, namely low mass
star formation. This would provide an additional source of energy
injection and generates flows at smaller scales of a few arcseconds,
1000-2000 AU, through bipolar outflows from isolated protostars or
multipolar outflows from multiple star formation (see
Figs.~\ref{complexareas}a-c).

%
\begin{figure*}
\resizebox{\hsize}{!}{\includegraphics{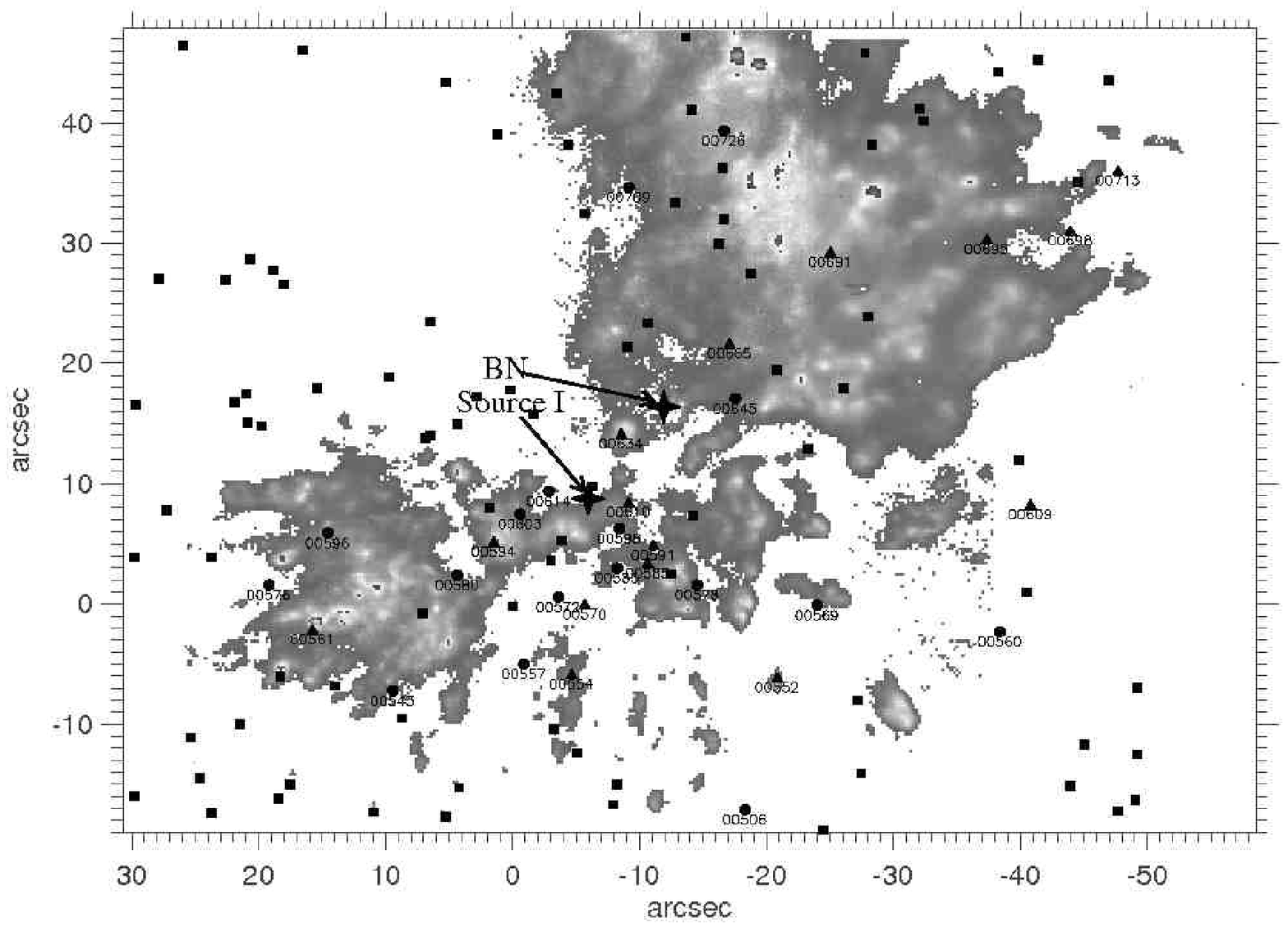}} 
\caption{\refe{The positions of point-like sources
    from \citet{muench02} in the observed field of OMC1. Objects have
    been divided into 3 categories based whether they are candidates
    (circles), possible candidates (triangles) or non-candidates
    (squares). Numbers refer to the labels in table~\ref{table_psc}.}}
\label{starplot}
\end{figure*}
%
\begin{figure*}
\resizebox{\hsize}{!}{\includegraphics{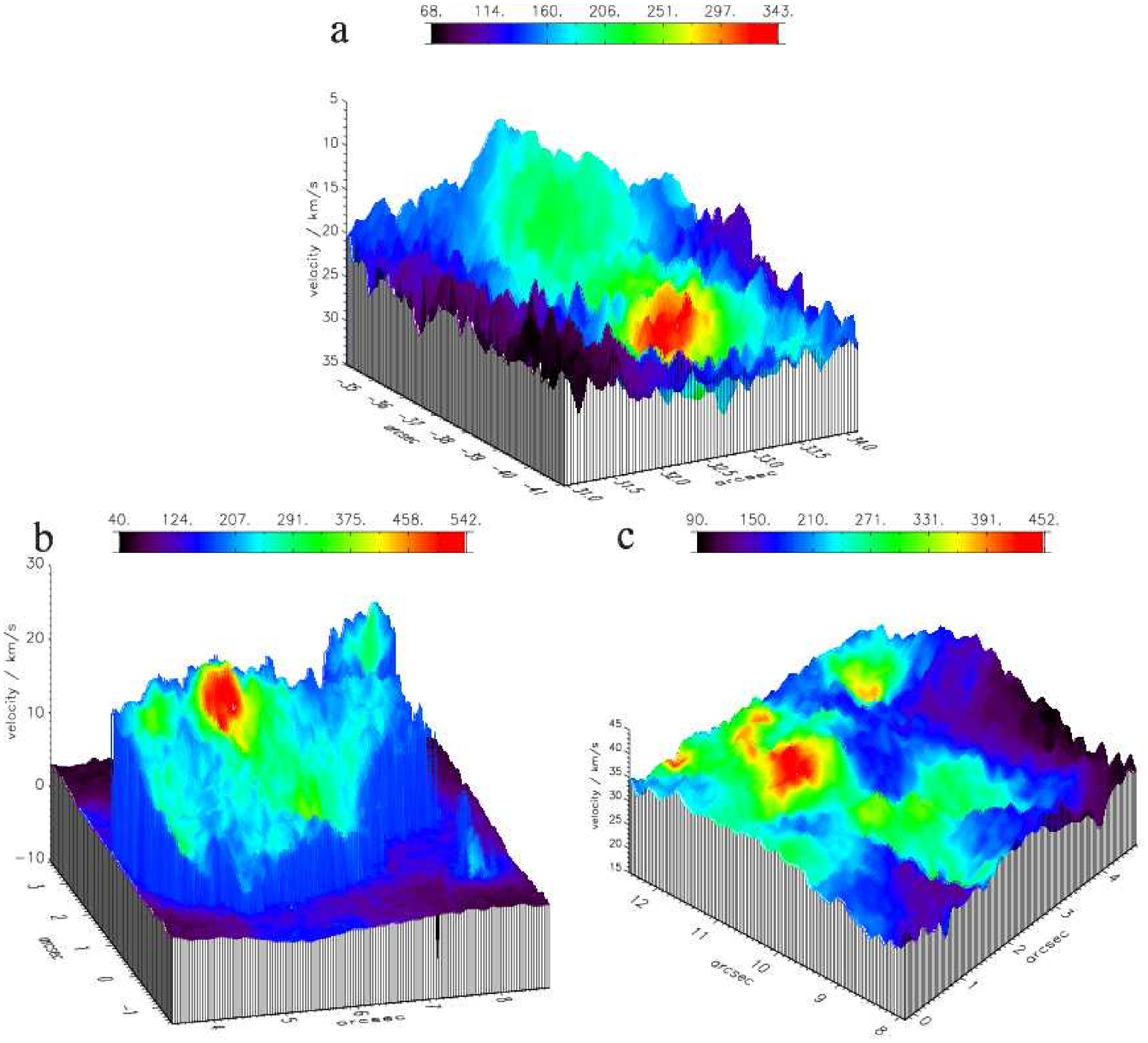}} 
\caption{Various areas with multiple flows in a small area: \textbf{a}
  A \refe{possible} bipolar outflow. Flows no. 1-76 and 1-81: see
  Table~\ref{table_p11}. \textbf{b} A complex flow pattern. Flows
  no. 2-42, 45, 46, 47, 48, 49: see Table~\ref{table_p2}. \textbf{c}
  Another example of a complex flow pattern. Flows no. 2-16, 18, 19,
  20, 24, 25: see Table~\ref{table_p2}. \refe{Colours denote the
  brightness in counts per 400~s exposure.}}
\label{complexareas}
\end{figure*}

\refe{We first consider the IR observational evidence for the presence
  of early protostars in OMC1. To this end we have compiled a table of
  protostellar candidates in OMC1 (table~\ref{table_psc}). We use the
  criterion reported in \citet{kenyon95}, for YSOs in Taurus, that an
  object with K$-$L $>$ 1.5 is almost always of a protostellar nature
  (specifically class I) and assume that this also holds for Orion. We
  apply this criterion to the photometry data of \citet{muench02} and
  \citet{lada04} for Orion. We furthermore use the criterion that
  J$-$H must be $>$1.0 to exclude, for example, unresolved binaries
  \citep{lada00}. The resulting protostellar candidates are shown in
  table~\ref{table_psc}. For ease of reference we have adopted the
  numbering system of \citet{muench02}. Where a ''$>$'' symbol is
  encountered, this implies that J band data are absent, in the upper
  section of the table, or H-band are absent in the lower.

We have divided objects into those which are very likely protostars,
labelled ''candidates'' (upper part of table~\ref{table_psc}) and
those where the identification is more tentative, labelled ''possible
candidates'' (lower half of table~\ref{table_psc}). In addition there
are objects for which the colours suggest that they may be excluded as
protostars. These are not shown in table~\ref{table_psc}. This last
category forms the majority of point-like objects within our complete
field. However within Peaks 1 and 2, these objects occur in roughly
equal numbers to the sum of ''candidates'' and ''possible
candidates''. Fig.~\ref{starplot} shows the positions of all objects
in the 3 categories. Candidates and possible candidates have been
marked with circles and triangles respectively with their accompanying
number for reference to table~\ref{table_psc}. Non-candidates have
been marked with a square. From a comparison with fig.~1 in
\citet{shuping04} we find that among the objects in
table~\ref{table_psc} are source n (00598), IRc4 (00585), IRc7
(00610), IRc11 (00614) IRc12 (00603) and IRc15 (00634). Furthermore
IRc9 is found to be identical with 00726. Source n is discussed in
more detail in section~\ref{sec:d-outflow}. IRc9 will be mentioned in
the following section, sect.~\ref{sec:d-edge}.}

\refe{The term
  ''candidates'' covers objects which have K$-$L $>$ 1.5 and which
  have either J$-$H $>$ 1.0 or where J$-$H is undetermined. The term
  ''possible candidates'' covers objects with H$-$K $>$ 2.0. As far as
  the authors are aware, the latter is a new criterion for the
  identification of protostars, motivated by a lack of L-band
  photometry. The new criterion has been established as follows. An
  extensive analysis of the H-K colour index of the stars in
  \citet{muench02} has been carried out. It has been found that for
  objects with K$-$L $>$ 1.5, 39\% have H$-$K $>$ 1.5 and 14\% have
  H$-$K $>$ 2. By contrast, for objects with K$-$L $<$ 1.5 -- which
  are not protostellar candidates -- less than 2\% have H$-$K $>$ 1.5
  and only one object (0.3\%) has H$-$K $>$ 2. On this basis a H$-$K
  colour index of $>$ 2 is adopted as a tentative identification of
  ''possible protostars'' in table~\ref{table_psc} and in
  Fig.~\ref{starplot}. We note that the value of H$-$K cannot however
  be used to exclude protostellar candidates. In this sense the
  objects listed in table~\ref{table_psc} form a minimum set. Under
  ''possible candidates'' in table~\ref{table_psc} we also list object
  from \citet{muench02} which have only been detected in K band. These
  are shown with H$-$K$>$~some value. The K-band detection, without J
  and H, suggests that these objects are deeply buried in the dusty
  gas. We conclude that there 17 good protostellar candidates in our
  field with a further 15 (or more) possible candidates. A comparison
  has also been made with the Chandra Orion Ultradeep Project (COUP)
  X-ray data~\citep{getman05,grosso05} and sources in
  table~\ref{table_psc}. COUP contains numerous point sources in the
  direction of the region studied here and a detailed comparison will
  be made elsewhere. Suffice it say at present that there is a
  remarkable correspondence between candidates in
  table~\ref{table_psc} with COUP sources with 14 out of 17 spatial
  overlaps, whilst the so-called possible candidates are
  anti-correlated, with only 3 out of 15 spatial overlaps. Overall
  about half of~\cite{muench02} sources are spatially coincident with
  COUP sources.}

\refe{We now turn to the positional correlation of these objects with
possible outflows and H$_2$ emission in general.} In some regions two
closely spaced shocks moving in opposite directions are seen, which
could indicate that a protostar is situated 
in between, causing the shocks. An example of this is shown in
Fig.~\ref{complexareas}a, where a pair of flows are found, one
red-shifted and the other blue-shifted. Data in table~\ref{table_p11}
show that position angles of these flows are roughly 180\degr\ apart,
noting the indeterminacy with regard to their relative
direction. \refe{However, if the positions of protostellar candidates
  in table~\ref{table_psc} are compared with the positions of
  candidate bipolar outflows, there is no clear correlation between
  the two. In only 2 positions, 15.75\arcsec~E -2.1\arcsec~N (object
  00561, see table~\ref{table_psc}) and 1.50\arcsec~E 5.2\arcsec~N
  (00594) a candidate protostar appears directly over a pair of flows,
  whereas in many cases a ''bipolar pair'' is found where there is no
  evidence of a protostar. Thus a ''bipolar pair'' in itself does not
  appear to be evidence of the presence of a protostar. However the
  caution should be made that we are considering the earliest stages
  of protostellar development (class 0/I). This phase is characterized
  by a very high degree of obscuration, and new, deeper observations
  than those of \citet{muench02}, used here to detect the protostars,
  may yield different results.}

As described in sect.~\ref{sec:f-grouping}, there is a marked
difference in how flows are distributed in Peak~1 and Peak~2. In
Peak~1 the flows are mostly uniformly distributed across the area,
whilst in Peak~2 the shocks often appear in small groups of 6-8 shocks
within an area of $\sim$5\arcsec\ in diameter (see
Fig.~\ref{density-blue-red} and Figs.~\ref{complexareas}b and c),
which corresponds to $\sim$2400~AU, at the distance of Orion. Since
most stars are expected to form as part of a binary or multiple system
\citep{pudritz02,larson03}, it seems plausible that these groups of
shocks indicate places where multiple star formation is ongoing: the
presence of 2-3 protostars would account for 4-6 shocks from bipolar
outflows. The flows in these regions are not however generally found
as pairs of shocks moving in opposite directions, as may be seen in
Figs.~\ref{complexareas}b and c with reference to the position angles
in Table~\ref{table_p2}. This supports a picture in which multiple
star formation is a complex process involving interactions between the
protostars and disks causing jets which are episodic or pulsed \citep[
  and references therein]{larson03}. 

\refe{We now examine whether we can associate regions with large
  densities of flows, as just mentioned, with protostars listed in
  table~\ref{table_psc}. Moving from east to west (left to right) in
  fig.~\ref{density-blue-red} the first region of high flow density is
  around coordinates 7\arcsec~E 0\arcsec~N. Only one stellar-like
  object is observed along the line-of-sight to this region: the
  proplyd 154-240 \citep{odell97c,schultz99}. Since this object has
  been observed in the visible by HST it cannot be a protostar buried
  in the dense, dusty gas.

Moving away from a line-of-sight positional correlation of possible
protostars and H$_2$ emission, there is a good protostellar 
candidate $\sim$4\arcsec~NW or 1600-2000~AU in the plane of the
sky. This is the object 00580 (\ref{table_psc}) which has K$-$L, J$-$H $>$
2 fulfilling our criteria for a buried protostar. This is
therefore a good candidate to power at least some of the emission that
is observed in this highly active zone. 

The second region of high flow activity in fig.~\ref{density-blue-red}
is around coordinates 0\arcsec~E 6\arcsec~N. Within 
a radius of a few arcseconds there is one star which satisfies the
protostellar criteria, namely 00603 with K-L = 2.22 and J-H =
1.02. There are two other stars for which we do not have L band data
available and therefore we are unsure of their nature. One of these is
00594 which is listed in table~\ref{table_psc} under ''possible
candidates''.  This object lies just outside the \citet{lada04}
frame. \citet{muench02} do not detect this object in J and H and
record m$_K$ = 13.95. Thus it is a highly obscured object and
therefore accordingly appears in table~\ref{table_psc}. A further
object is 00606. This object has no L-band data, and with H$-$K = 0.57
this is not an obvious protostellar candidate. However, one should
recall from the above discussion that the H-K colour index is useful
only for including, not excluding, protostellar candidates. 

We turn now to the third flow concentration, which is centred on
$-$22\arcsec~E 37\arcsec~N. Surveying the region within a radius of
$\sim$ 4\arcsec, we find the star IRc9 (00726)~\citep{smith05}, with
J-H and K-L both $>$2, fulfilling the protstellar criteria, see
table~\ref{table_psc}. There is also another object (00715) within a
few arcseconds from this concentration of flows, but again there is
insufficient photometry to make an assignment as to the nature of this
object.

Looking at Peak 1 in general, rather than at concentrations of flows,
there are a further two protostellar candidates, 00645 and 00709A, both
situated along the edge of the H$_2$ emission seen in Peak~1. There
are also another 5 possible protostellar candidates (00665, 00691,
00695, 00698, 00713) which are spread across Peak~1. In Peak 2, there
are a further 3 protostellar candidates (00543, 00576, 00596), and one
possible candidate, 00561, associated with the 
very bright H$_2$ emission at 16\arcsec~E -2\arcsec~N. The object 00561 has
in fact an associated 140~AU diameter object observed in the L-band
with the VLT (Paper I), as mentioned in sect.~\ref{sec:shockPDR}.
We note that with the exception of 00561 (in Peak 2) no stars or
protostars in Peak~1 or 2 are directly associated with the brightest
areas of H$_2$ emission. 

In region B there are 8 protostellar candidates and a further 8
possible candidates. Half of these are situated in the dense core of
OMC1 around Source~I and may contribute to H$_2$ emission in this
region. The other half are widely distributed over region B.} 

The discussion here and in preceding sections points towards Peak~2
being a more active star forming region than Peak~1 \refe{with similar
numbers of protostellar candidates despite Peak~2 being only half the
size of Peak~1 in the plane of the sky}. This is an
unexpected result given the proximity of the two regions. One possible
explanation is that the ionization front from \thetc\ has cleared away
most of the diffuse gas in Peak~2, leaving only the densest parts of a
previously more extensive cloud and allowing us to see the dense star
forming cores which would otherwise have been obscured from view. This
interpretation is supported by the work of \citet{pogge92} and
\citet{takami02} who show evidence of an ionization front just north
of Peak~2. 

The above considerations ignored the presence of WRVs in Peaks~1
and~2. In Peak 2 WRVs form only 20\% of the observed flows but in Peak
1, they form more than 50\% of the flows. However analysis presented
in sect.~\ref{sec:f-edge}, showing the organised nature of WRV flows in
Peak 1, suggests that WRVs form a subset of flows which should not be
associated with local star formation. This is further discussed in the
section directly below, sect.~\ref{sec:d-edge}.  

\refe{At the start of this section we asked whether H$_2$ emission
  might in part be caused by the presence of protostars with
  associated outflows. Our data show that this is the case but we do
  not demonstrate that the majority of bright emission is due to
  protostellar activity in OMC1. By their nature the most important
  sources may be deeply buried. However for the present we conclude
  that the emission arises from a mix of protostellar and massive star
  activity. In particular in Peak 1, present data suggest that
  emission may be dominated by massive outflow from the vicinity of
  Source~I and the BN region.}

\subsection{The origin of weak radial velocity structure in Peak 1} 
\label{sec:d-edge}
The spatial distribution of weak radial velocity components (WRVs) in
Peak 1 revealed a band, $\sim$2--3\arcsec\ in width and more
than 30\arcsec\ long passing diagonally across emission in Peak 1 (see
sect.~\ref{sec:f-edge}). It was also noted that 2/3 of these WRVs showed
a much more broken-up structure than other features, including other
WRVs. This section briefly explores possible scenarios for the
origin of this band of WRVs. 

The first scenario is that the band of WRVs may represent a less
energetic part of the outburst which created the ''bullets'' or
''fingers''~\citep{allen93}. These are clumps of gas moving radially
outwards at high velocities of up to several hundred\kms\
\citep{doi02} and appear to have an origin in the area around BN-IRc2
and Source~I \citep{salas99,odell01,doi02}. If the WRVs are moving at a
velocity of 40\kms\ -- the upper limit of motions observed here for
other clumps of H$_2$ -- in 1000 years they would be found at a
position about 18\arcsec\ from the BN-IRc2 complex, essentially where
they are observed. The typical mass of each WRV can be estimated to be
$\sim$~1.5\,$\times$\,10$^{-3}$\,M$_\odot$, given a density of
10$^7$\,cm$^{-3}$ and a diameter of 500\,AU. This yields a total bulk
energy for outflowing hot H$_2$ in the WRVs of the order of
$\sim$~5\,$\times$\,10$^{44}$~ergs, for 20 such objects. This is about
1\% of the energy associated with the bullets whose properties are
reported in \citet{burton97}.

A second scenario is that the band of WRVs is the result of
bright photodissociation zones powered
by \thetc, but with enhanced brightness due to geometrical effects
(see below). The morphology of the ionization front, described in
\citet{wen95} and refined in \citet{odell01}, shows that the bright
H$_2$ clumps observed in the present work arise from dense gas  
within the \hii\ region created by \thetc\ \citep{k2003}. Ionizing
radiation has cleared away the diffuse gas and is now slowly eating
its way through the denser clumps of gas remaining. Thus the line of
WRVs along the eastern edge of Peak 1 may be the relic of a denser
region that existed in the original molecular
cloud. Whilst the brightness of the H$_2$ emission cannot be
reconciled with a face-on PDR (see sect.~\ref{sec:shockPDR}), PDR 
emission may be enhanced locally through a geometry originally
proposed in \citet{field94}. In this model, a choppy surface of H$_2$
yields chance lines-of-sight with high column density. This is
consistent with the broken up appearance of 2/3 of the features in
this zone, mentioned in sect.~\ref{sec:f-edge}.

\refe{If we consider only those WRVs lying in the band of width 2.5\arcsec\,
between points P and Q in fig.~\ref{intplot}, a further possibility is
that these may be associated with an outflow in the plane of the sky
from IRc9 (00726 in table~\ref{table_psc} and fig.~\ref{starplot}). IRc9
is located ~$\sim$2\arcsec\ from the centre of this
band. \cite{smith05} show that IRc9 is a YSO possessing a clearly
defined disk. The major axis of the disk is oriented at right-angles
to the WRV band which suggests that WRVs in the band could represent a
highly collimated outflow in the plane of the sky from IRc9. 
In sect.~\ref{sec:f-edge} we found that there was a significant
overrepresentation of WRVs in Peak~1. If the analysis is repeated but
without including the WRVs in the band, a significant
overrepresentation of WRVs in Peak~1 is still found: based on the same
model used in sect.~\ref{sec:f-edge} the expected number of WRVs in
Peak~1 is $\sim$12 while 32 WRVs are actually observed. Thus a
significant discrepancy remains.}

\section{The outflow zone, Region B} \label{sec:d-outflow}
\subsection{The geometry of the region: radio observations and the
  present data} \label{sec:geometry} 
\begin{figure}
\begin{center}
\resizebox{\hsize}{!}{\includegraphics{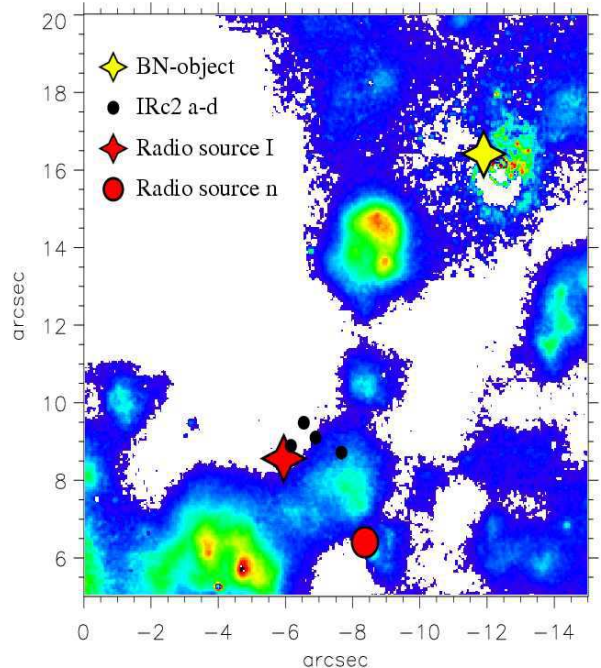}}
\caption{The positions of BN, radio source I, radio source n and
  IRc2 a-d, superimposed on velocity integrated emission in
  the v=1-0 H$_2$ S(1) line.} 
\label{HCmap}
\end{center}
\end{figure}

An outflow in Region~B has previously been identified in
radio-observations involving both proper motions and radial velocity
measurements of H$_2$O, OH and SiO masers and thermal emission in SiO
(v=0) and NH$_3$. Evidence of shocked gas within the flow is provided
through observations of emission from high rotational states of SO,
SO$_2$, HCN, where data are reviewed in \citet{genzel89}. The outflow
has also been identified in subsequent observations reported in
\citet{wright92,wright96} in data involving 16 molecular species, in
particular SO and SO$_2$ \citep{wright96}. The outflow is 
associated with source I whose location is given in Fig.~\ref{HCmap}. 

The expansion velocity associated with the above-mentioned outflow in
Region~B on the basis of radio observations was found to be
18$\pm$2\kms\ \citep{genzel81}. The data of \citet{genzel81} showed 
a flow extending $\sim$20\arcsec\ SW of the BN-IRc2 region, which is
very similar to the SW extent of the H$_2$ emission recorded
here. \citet{genzel81} also note, through their observations of
numerous objects at closely similar velocity, that the flow appears to
have been stable for $\gtrsim$3000 years. This would seem to preclude
the scenarios discussed in \citet{shuping04} in which source~I may be
the origin of both of the major outflows in this region, through
unstable behaviour, such as flipping of the orientation of the
outflow. 
%
%

\citet{menten95} performed SiO maser observations of the centre of
the region around source I, extending over $\sim$50\,AU (0.11\arcsec),
and found both blue- and red-shifted flows, with the blue-shifted
component showing a v$_{lsr}$ between $-13.0$ and $-6.2$\kms\
corresponding to a blue-shifted flow relative to 
the ambient gas of between 25 and 18 $\pm$6\kms, taking into account the
uncertainty in the v$_{lsr}$. The centre of the SiO maser emission was
found to be located at the position of source I. The same spatial
relationship of SiO masers to source I is recorded in
\citet{doeleman99}. Further recent SiO observations covered an
inner area of $\leq$2\arcsec\ diameter (900 AU) and involved both SiO
maser and non-maser emission, the latter in v=0, and maser emission in
H$_2$O, including proper motion studies \citep{greenhill04a}. As in
\citet{menten95}, the expansion centre of these motions was found to
be coincident with source I, in this case to within 30 mas ($\sim$15
AU). Radio source n (hereafter simply source n, see
Fig.~\ref{HCmap}) a few arcseconds from source 
I is also significant. From the observations of \citet{genzel81} and
\citet{menten95} source n is found at the centre of expansion of a
shell of H$_2$O masers, perhaps constituting an independent outflow
zone~\citep{shuping04}.

Our present observations measure gas motions relative to the ambient
surrounding gas and therefore may also be interpreted as a radial
component of an expansion velocity. The expansion velocity of
18$\pm$2\kms\ derived from the observations of \citet{genzel81}
matches accurately the mean velocity of $-18$\kms\ observed here for
the hot H$_2$ in Region~B: see Fig.~\ref{stat-vel}. The velocities
measured in this work are also consistent with data in
\citet{menten95} and \citet{wright96}. 

The much debated issue of the origin of large scale outflows in OMC1,
examined in detail for example in \citet{menten95} and elsewhere, has
recently achieved a new perspective with high spatial resolution
(0.3--0.5\arcsec) IR measurements in \citet{shuping04} and
\citet{greenhill04b}. These observations show that there are a number
of potential YSO candidates buried within a zone in the north-east of
Region B. IRc2 was long believed to be the main power source in OMC1
and the source of the outflow activity in the region. However these
and other high resolution observations \citep{Dougados93} have
resolved IRc2 into several individual sources. These may or may not be
self-luminous \citep{shuping04}. At all events the current most likely
candidates for outflow sources are sources I and n.

%
\begin{figure*}
\resizebox{\hsize}{!}{\includegraphics{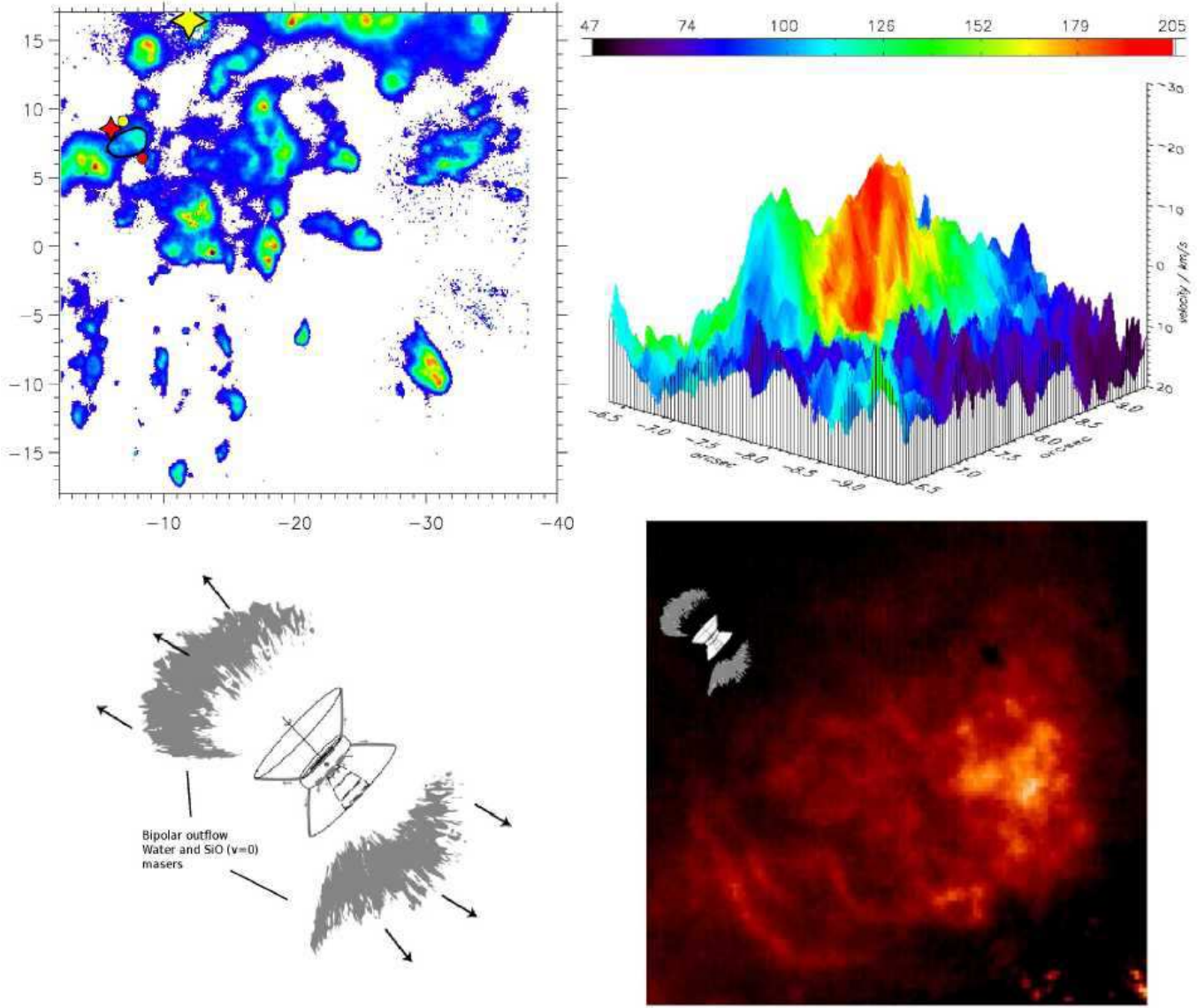}} 
\caption{Source~I. \textbf{Top left:} Region B recorded in the v=1-0
  S(1) line of H$_2$. The yellow star shows the position of BN and the
  red star shows the position of source~I. The red circle is source n,
  and the yellow circle is IRc2 d (compare to Fig.~\ref{HCmap}). The
  black outline shows the position of the H$_2$ emission feature shown
  in the right hand panels. \textbf{Bottom left:} the geometry of the
  wide-angle outflow from source~I taken from
  \citet{greenhill04a}. \textbf{Bottom right:} VLT NACO image
  \citep{lacombe04} of the outflow region (the ''fireball'') just SW
  of source~I, outlined in black in the top left panel. The inset
  shows that the outflow from source~I (not to scale) points directly
  at this region. \textbf{Top right:} velocity structure of the
  fireball, showing three fast jets impinging on the medium (flows
  no. B-7, B-9, B-10: see Table~\ref{table_cz}).} 
\label{sourcei}
\end{figure*}
%
The positions of sources are shown in Fig.~\ref{HCmap}. Source~I
($-$5.95\arcsec~E, 8.56\arcsec~N), is a deeply buried O-star
\citep{menten95,greenhill04a}. It is the most luminous source in the
hot core of OMC1 with a luminosity of 5$\times$10$^3$ to 10$^4$\,L$_\odot$
\citep[ and see sect.~\ref{sec:properties}]{menten95,beuther04}, compared for
example to $\sim$2500\,L$_\odot$ for the BN-object \citep {gezari98}
or $\sim$2000\,L$_\odot$ estimated for source~n \citep{greenhill04b}.
Source I is obscured in the infrared and cannot be detected at
wavelengths as long as 22~$\mu$m~\citep{greenhill04b}. Source n
\citep{greenhill04b} is located at $-$8.35\arcsec~E, 6.40\arcsec~N and
according to \citet{shuping04} may give rise to an outflow in this
region (see above and sect.~\ref{sec:sourcen}).

Based on their data for SiO and H$_2$O emission around source~I,
\citet{greenhill04a} proposed a model with an accretion disk and a 
wide angle outflow pointing NE-SW with an opening angle of
$\sim$90\degr. This model replaces that of \citet{greenhill98}. The
morphology of the outflow from source I, newly proposed in
\citet{greenhill04a}, is shown in Fig.~\ref{sourcei}, lower left
panel. 

The axis of the outflow from source I passes through a region where we
find a striking H$_2$ emission feature. Velocity data for this zone is shown
in the upper right hand frame of Fig.~\ref{sourcei}. Three fast jets
(flows no. B-7, B-9, B-10: see Table~\ref{table_cz}) are seen
impinging on the surrounding medium. A VLT image of the same area
\citep{lacombe04} is shown in the lower right-hand frame of
Fig.~\ref{sourcei}. The model of the outflow from source I is
superimposed upon it, to show the spatial relationship, where the
accuracy of the relative positions is $\sim$0.2\arcsec. The position
of the maximum brightness of this outflow region, measured in H$_2$,
is 1.84\arcsec~W and 0.96\arcsec~S of source I. The appearance of the
H$_2$ emission in this region suggests that this is a zone where the
radio-detected outflow bursts through the dense gas surrounding
source~I. Thus our observations show the IR counterpart of the O-star
outflow, hitherto only observed at radio-wavelengths. 

\begin{figure*}
\center
\resizebox{0.8\hsize}{!}{\includegraphics{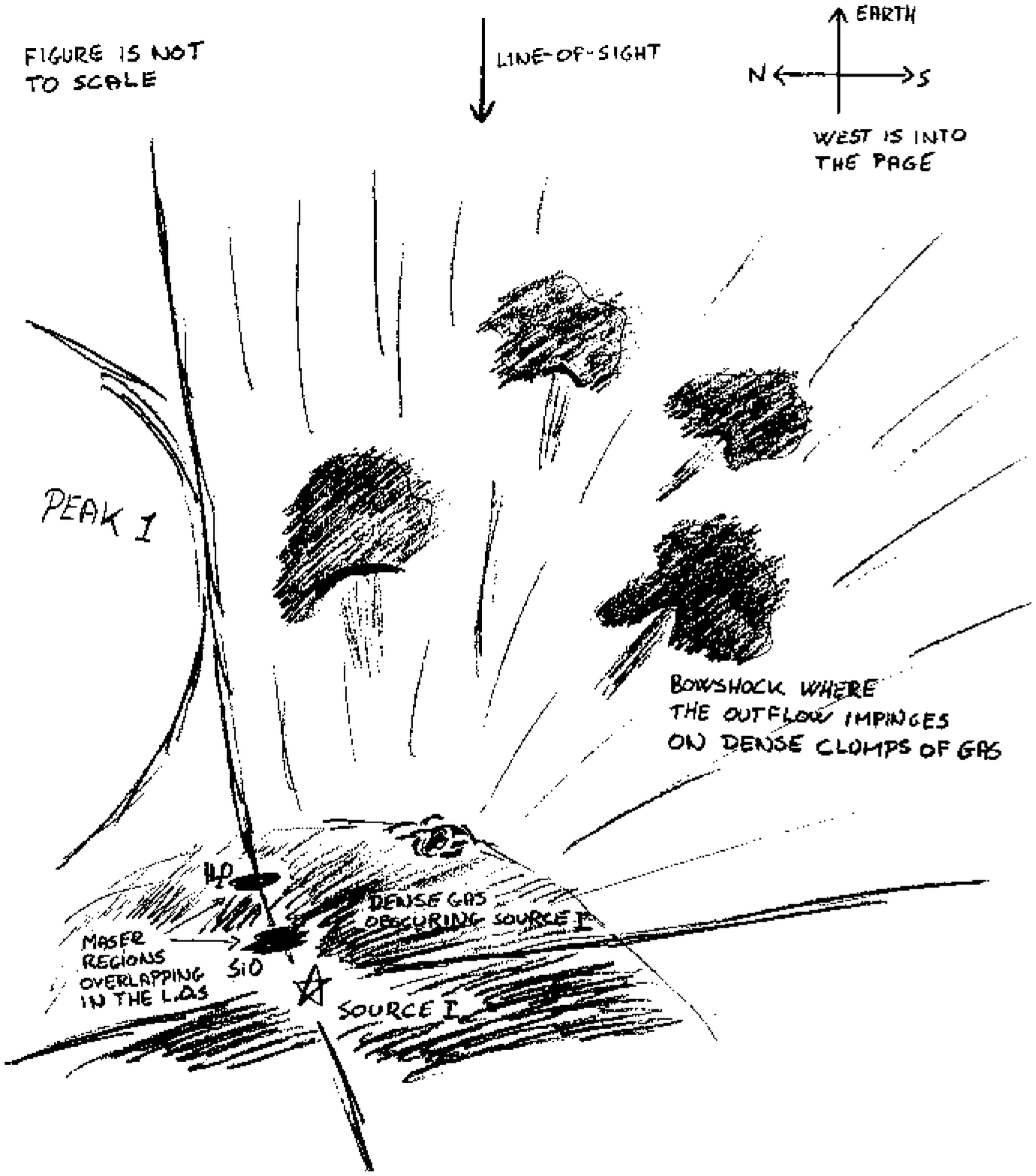}} 
\caption{The outflow from source I, showing the appearance of the
  outflow and its relationship to other objects in Region B, seen at
  right-angles to the line-of-sight from a vantage point in Peak~2
  looking towards the west.}
\label{drawing}
\end{figure*}
%
The model of \citet{greenhill04a} of the outflow from Source I is of an
O-star with an edge-on disk and a wide-angle (90\degr) bipolar outflow in the
plane of the sky. This begs the question of why no H$_2$
emission is observed as part of the outflow to the NE of
source I. We would suggest that the flow axis may in fact tip out of
the plane of the sky by $\lesssim$45\degr. Flows would
then tend to be rendered invisible though dust obscuration in this
very dense zone. Moreover, \citet{greenhill04a} observe a substantial
overlap of H$_2$O and SiO masers along the line-of-sight. However
these two molecular masers trace different gas densities and
temperatures, and as \citet{greenhill04a} note should not be found
within the same mass of gas. H$_2$O masers may have associated kinetic
temperature of 200 to 2000\,K, with a preference for temperatures
around 600\,K, and densities of no more than a
few\,$\times$\,10$^9$\cm~but most likely with densities $<$10$^9$\cm\
\citep{yates97}. SiO masers require SiO to remain in the gas phase,
with associated temperatures of $>$1500K and densities of
10$^{10}$\cm\ \citep{doel95,humphreys96}. If, however, the outflow
cone has a significant component in the line-of-sight, as we suggest
here, this discrepancy may be resolved: the H$_2$O masers lie further
from the O-star than the SiO masers, but overlap in the line of sight. 
Further, blue-shifted flows are observed to the north of Source
I. This is consistent with an outflow originating from source I, but
tipped out of the plane as we suggest. This is illustrated in
Fig.~\ref{drawing} -- described in more detail below -- where any
material to the left of the line-of-sight through source I would
effectively appear as blue-shifted material to the north of source I. 

In conclusion we agree with \citet{greenhill04a} that the direction of
the outflow in the plane of the sky is NE-SW. Following the result of
sect.~\ref{sec:orient}, the position angle of the axis of the cone of
the outflow from source~I is $\sim$240\degr$\pm$10\degr. We suggest
however that the flow, rather than lying in the plane of the sky, lies
at an angle of as much as 45\degr\ to this plane.

A schematic diagram of the region, not to scale, based upon this
model, may be found in Fig.~\ref{drawing}. This shows the region as if
observed from right-angles to the line-of-sight, from a position in
Peak~2, looking west. Starting from source I at the base of the
diagram, first SiO and then H$_2$O masing regions, overlapping in the
line-of-sight, are found in the dense gas close to source I. Moving
further from the O-star, the outflow breaks through the dense gas and
dust surrounding source I, generating H$_2$ emission features such as
that shown in Fig.~\ref{sourcei} and, moving still further, other
blue-shifted features shown in Fig.~\ref{intplot} and listed in
Table~\ref{table_cz}.

The model proposed above requires that the
directions of flows of H$_2$ emitting regions should show asymmetry in
the NE-SW direction. Referring to Fig.~\ref{intplot} and
Fig.~\ref{anisotropy}, the star diagram for Region~B in the upper left
panel shows a 3:1 anisotropy about a dividing line running roughly
perpendicular to the outflow direction. In this connection, recollect
that the absolute direction of the flow, $\pm$180\degr\, is not
determined, as noted earlier. The only additional assumption
at this juncture is that one or other model on which the flow
direction is based is dominant, see sect.~\ref{sec:orient}. 

\subsection{Physical properties of the outflow source}
\label{sec:properties}
This section considers some of the governing parameters of the
O-star associated with source I and with
region~B. \citet{greenhill04b} found that source~I could not be 
detected at 22\,$\mu$m. They concluded that the optical depth at
22\,$\mu$m must be $>$300. Using the relationship between extinction
and wavelength given in \citet{rosenthal00}, the NIR optical depth
will then be $>$570. This implies that the column density of H$_2$ is 
$>1.5$\,$\times$\,10$^{25}$~cm$^{-2}$. This is comparable with the
figure of $\leq$8.5\,$\times$\,10$^{24}$~cm$^{-2}$ estimated in
\citet{beuther04}. Depending on the average gas density around the
obscured O-star, this object must be buried several arcseconds within
OMC1. For example if the average density is 10$^9$\cm, the O-star is
buried to a depth of 2\arcsec\ and the mass of gas surrounding the
O-star is 9~M$_{\odot}$. This is comparable with the estimate of the
mass of the O-star itself of $\leq$10~M$_{\odot}$ in \citet{beuther04}
or the figure of 10-20~M$_{\odot}$ given in \citet{shuping04}. These
masses are consistent with a luminosity of a few thousand to
$\sim$~10$^4$~L$_{\odot}$, consistent also with \citet{menten95}. The
most recent estimate places the luminosity of source I at
$\sim$5$\times$10$^3$~L$_{\odot}$ \citep{beuther06,beuther04}.

The mass outflow rate in Region~B may be estimated as follows. Typical gas
densities in the H$_2$ clumps in Region~B have been investigated in
detail in \citet{k2006} and post-shock values range between 10$^6$ to
10$^7$\cm, equivalent to an average density of
$\sim$10$^{-17}$~g\,cm$^{-3}$. Assuming a continuous outflow and a measured
velocity of 18\kms, this represents a mass flux of
2\,$\times$\,10$^{-11}$~g\,cm$^{-2}$\,s$^{-1}$. The total area in the
plane of the sky covered by the excited H$_2$ may estimated by using a
total area for each clump given by the area around each flow where the
emission is $\geq$50\% of the maximum brightness of the flow. The
total area is found to be $\sim$5\,$\times$\,10$^{33}$~cm$^2$, giving
a mass flow of
$\sim$1.5\,$\times$\,10$^{-3}$~M$_\odot$\,yr$^{-1}$. Including the 
red-shifted part of a bipolar outflow, this gives a total outflow rate
of $\sim$3\,$\times$\,10$^{-3}$~M$_\odot$\,yr$^{-1}$. Since only
excited gas has been included this is a lower limit. This figure
for the outflow rate is comparable to estimates given
in \citet{shepherd05} for stars of the luminosity 
associated with source I.

An independent estimate of the outflow rate may also be performed. If the
luminosity of the source is accretion luminosity and the surface
temperature is (say) 1500~K, typical of a protostar, then the stellar
radius is $\sim$5~AU, given the figure of 5$ \times
10^3$~L$_{\odot}$. The mass infall rate is then
0.02~M$_\odot$\,yr$^{-1}$. For low mass stars the outflow rate is
generally $\sim$10\% of the infall rate \citep{richer00}. If the same
proportion is adopted for the case of  this massive star, the result
is a mass outflow rate which is essentially the same as that which has
been estimated from our observations. The corresponding outflow
momentum rate is $\sim$0.05~M$_\odot$\,yr$^{-1}$\kms\ and mechanical
luminosity is $\sim$80~L$_\odot$. These values are typical of a
late-type O-star or early-type B-star \citep{shepherd05}. Together
these characteristic properties combine to make a consistent picture
of an outflow from source I where this source is a massive and very
young star.

\subsection{Source n} \label{sec:sourcen}
As already noted, source n \citep{menten95} is another major power
source in the region, in addition to source I and BN. Source I and n
lie on a line running NE-SW as shown in fig.~\ref{HCmap}. Therefore 
source n is also located on the axis of the low-velocity outflow in
Region~B. Furthermore \citet{shuping04} and \citet{greenhill04b}, see
also \citet{beuther04}, have made observations which suggest the
presence of a disk around source n with the rotation axis aligned with
the low-velocity outflow. Thus source n, which is a possible centre of
expansion of numerous H$_2$O masers in the vicinity
\citep{genzel81,menten95}, may be the source of or may 
contribute to the low-velocity outflow in Region~B. However, at
present source I would seem the clearer candidate following the
observations of \citet{greenhill04a} and also the fact that source n
may have a relatively low luminosity of $\sim$2000~L$_\odot$
\citep{greenhill04b} compared to source~I. 
%
%
%
\section{Conclusions}

The major conclusions of this study using GriF data for the central
region of Orion OMC1 may be summarised as follows.\\
\\
\emph{(i)} 193 distinct bright H$_2$ emission features in the
\mbox{v=1-0 S(1)} line 
have been identified. 67\% of these show clear radial velocity
structure with well defined features of greater than 5\kms, indicating
that the emission is generated by shocks in the medium. \\
\\
\emph{(ii)} There are equal numbers of flows in Peaks 1 and 2, implying a
greater concentration of activity in Peak 2 which is less than half
the size of Peak 1. Flows are found to be grouped much more tightly in
Peak 2 than in Peak 1. Blue shifted flows are somewhat favoured over 
red-shifted, presumably due to dust obscuration. No velocities exceed
35--40\kms, consistent with the upper limit for the propagation of
C-type shocks in dense media. \\
\\
\emph{(iii)} In a distinct zone, Region~B, lying south of Peak 1 and west of
peak 2, all flows are blue-shifted, with an average velocity of
$-$18$\pm$8\kms. \\
\\
\emph{(iv)} Weak radial velocity features (WRVs) with radial velocities
$<$5\kms\ are strongly concentrated in Peak 1, and a band of such
structures has been identified in a NE-SW direction across Peak 1. In
this band, 2/3 of the structures show a broken-up morphology, not seen
elsewhere, which may arise from a major photodissociation contribution
to the emission through geometrical projection effects. This band of
structures may also be a further manifestation of an outburst from the
BN-IRc2 region about 1000 years ago which created the bullets or
fingers lying further to the NW. \\
\\
\emph{(v)} Spatial correlation between maximum brightness of H$_2$ emission and
maximum velocity reveals deviations typically of $\leq$0.6\arcsec. These
deviations may be used to estimate the orientation of flows, or
shocks, in the plane of the sky - but not their absolute direction,
that is, whether they are travelling north or south, say. Orientations
are found to be random in Peak~2 and with some clear preference for a
SE-NW direction in Peak~1, but show a strong preference for a NE-SW
orientation in Region~B.  \\ 
\\
\emph{(vi)} In Paper I it was noted that potential low mass star forming
zones, identified through the presence of energetic flows in OMC1,
showed higher densities and correspondingly higher energy flux, and
are altogether less simply structured than more conventional Class 0
or 1 low mass isolated star formation regions. Here we identify a
large number of zones which show similar characteristics to those
discussed in Paper 1. Data point in particular to the close physical
association of numerous strong H$_2$ emission features. It would seem
likely that these associations represent sites of multiple star
formation characterised by episodic outbursts of activity creating
several outflows in different directions.\\ 
\\
\emph{(vii)} The most striking feature of the results is the blue-shifted
outflow in Region~B. This is the infrared counterpart to the outflow
associated with the very young O-star, source I, a radio source with
associated SiO and H$_2$O masers and non-maser emission in numerous
molecules. Radio data show the presence of a blue-shifted flow, part
of a general expansion with an average velocity of 18 $\pm$6\kms, in
agreement with our value of $-18\pm$8\kms. \\
\\
\emph{(viii)} There is a remarkable feature in the H$_2$ emission data with
strong velocity components. On the basis of a VLT image, this is
suggestive of a region in which a flow has burst through the material
of OMC1 from behind, originating in source I and sending shock waves
through the medium. The axis of the radio outflow from source I passes
through this object. These and other data lead us to conclude that
source I is a major progenitor of activity in OMC1 and in particular
of the outflow in region~B.\\ 
\\
\emph{(ix)} There remain several extraordinary objects in the field whose
nature we do not comprehend. For example there is a structure,
3\arcsec~$\times$~3\arcsec, Table~\ref{table_p2}, object 2-32 and 2-34
in Peak 2, which is the only clump of emission in the entire field
with v$_{lsr}$ consistently around $>$40\kms\ \citep[shown in Fig.~4
  of][]{g2006a}. The origin of such an energetic object is unknown.\\
\\
In conclusion, the GriF data with its associated high spatial and
velocity resolution provide information which opens a new window on
OMC1 and presents a new perspective on the star-forming activity for
both massive and low mass stars. Data have very recently been obtained
for OMC1 using the VLT with adaptive optics and Fabry-Perot
interferometry in the v=1-0 S(1), S(0) and the v=2-1 S(1) lines. Shock
structures seen in these 3 lines will constrain the physical
conditions strongly in star forming zones in OMC1. The chief lack at
present is of shock models which contain both detailed chemistry, essential
for C-type shocks, and also the geometry, for example of bow
shocks. Models are under development and will be necessary to settle
such basic questions as the relative physical disposition of the
location of the maximum brightness and maximum velocity in H$_2$
emission features.

%
%

\begin{acknowledgements}
HDN, MG and DF would like to acknowledge the
support of the Aarhus Centre for Atomic Physics (ACAP), funded by the
Danish Basic Research Foundation. HDN and MG would also like to
acknowledge financial support from the Instrument Centre for Danish
Astrophysics (IDA), funded by the Danish National Science Committee
(SNF, now FNU). JLL would like to acknowledge the support of the PCMI National
Program, funded by the French Centre National de la Recherche
Scientifique (CNRS). We also wish to thank the Directors and Staff of
CFHT and of the VLT for making possible observations reported in this
paper and for the assistance rendered by E. Le Coarer (Observatoire de
Grenoble) in obtaining GriF data. We also wish to thank G. Pineau des
 For\^ets and Lars E. Kristensen for valuable discussions about the nature
of shocks. Our thanks are also due to the referee for many helpful
comments, and to whom we owe, among other items, the suggestion that
we include details of protostellar candidates and the model that the
outflow from IRc9 may be the source of some of the anomalous objects
found in our field.
\end{acknowledgements}
%
%
\bibliographystyle{aa}
\bibliography{bibliography}
\clearpage
%
%
\appendix
\section{Tables}
\begin{longtable}[h]{c c c c c c c c c c}
\caption{Features identified in OMC1 in Peak 1. Data are
  sorted by the east-coordinate of the maximum in brightness. Column
  \textbf{1} labels each feature and columns \textbf{2--5} show the
  position in the plane of the sky of the maxima in brightness and
  velocity for each of the features. Positions are measured in
  arcseconds east and north of the star TCC0016 (05$^{\rm h}$35$^{\rm
  m}$14\fs91, $-$05\degr22\arcmin39\farcs31 (J2000) ). Where no
  coordinates are given for a maximum in velocity, no well-defined
  peak in the velocity was found nearby. Columns \textbf{6} shows
  the velocity of the gas surrounding each feature expressed as
  v$_{lsr}$. Column \textbf{7} shows the maximum velocity of the flow
  relative to this local background velocity (see
  sect.~\ref{sec:findflows}). Column \textbf{8} contains the maximum
  brightness of each feature per 400~s exposure. Column \textbf{9}
  shows the position angle for each flow, measured east of north
  (i.e. counter-clockwise), see sect.~\ref{sec:sfrs}. A dash indicates
  that the object is a weak radial velocity region (WRV) and 'no p.a.'
  indicates a displacement between maximum velocity and brightness of
  $<$0.25\arcsec.  A star ($^\star$) marks flows where the 
  morphology of the emission clearly indicates the absolute direction
  of the flow (see sect.~\ref{sec:orient}): these are only found in
  Table~\ref{table_cz}. Column \textbf{10} gives the 
  displacement in the plane of the sky between the observed maxima in
  brightness and velocity for each feature. A summary of the data can
  be seen in Fig.~\ref{intplot}.}
\label{table_p11}\\
%
%
%
\multicolumn{10}{c}{Flows in Peak~1} \\
\hline \hline
 & \multicolumn{4}{c}{Coordinates of maxima} & velocity of &
flow & maximum & position & displace- \\
No. & \multicolumn{2}{c}{Brightness} & \multicolumn{2}{c}{Velocity} &   
surroundings & velocity & brightness & angle & ment \\
\hline

 & E & N & E & N & km\,s$^{-1}$ & km\,s$^{-1}$ & counts  &
degrees & arcsec \\
\hline
\endfirsthead
\caption{continued.}\\
\hline
 & \multicolumn{4}{c}{Coordinates of maxima} & velocity of & flow & maximum &
position & displace- \\
No. & \multicolumn{2}{c}{Brightness} & \multicolumn{2}{c}{Velocity} &
surroundings & velocity & brightness & angle & ment \\
\hline 
 & E & N & E & N & km\,s$^{-1}$ & km\,s$^{-1}$ & counts  &
degrees & arcsec \\
\hline
\endhead
%
\hline
\multicolumn{10}{r}{\emph{Continued on next page}}
\endfoot
\hline \hline
\endlastfoot
 1-1&	-3.61&	43.58&	--&	--&	17&	--&	204&	--&	--\\
 1-2&	-7.14&	43.68&	-7.11&	43.12&	16&	-6&	207&	356&	0.6\\
 1-3&	-7.77&	22.61&	-7.60&	22.54&	8&	-14&	421&	no p.a.&0.2\\
 1-4&	-8.51&	46.87&	-8.47&	47.39&	20&	9&	398&	184&	0.5\\
 1-5&	-9.48&	47.28&	--&	--&	20&	--&	368&	--&	--\\
 1-6&	-9.52&	27.09&	--&	--&	8&	--&	302&	--&	--\\
 1-7&	-9.90&	44.94&	--&	--&	21&	--&	302&	--&	--\\
 1-8&	-10.22&	42.46&	-10.05&	41.90&	21&	7&	264&	343&	0.6\\
 1-9&	-10.26&	46.38&	--&	--&	20&	--&	366&	--&	--\\
1-10&	-12.18&	29.89&	-13.06&	30.73&	13&	-16&	227&	134&	1.2\\
1-11&	-12.36&	36.85&	--&	--&	16&	--&	224&	--&	--\\
1-12&	-12.39&	28.59&	-11.94&	28.77&	13&	19&	196&	249&	0.5\\
1-13&	-13.16&	41.16&	--&	--&	17&	--&	290&	--&	--\\
1-14&	-13.61&	29.05&	--&	--&	13&	--&	229&	--&	--\\
1-15&	-13.86&	22.61&	-13.48&	22.43&	6&	-28&	187&	294&	0.4\\
1-16&	-14.18&	40.95&	--&	--&	18&	--&	328&	--&	--\\
1-17&	-14.91&	46.72&	--&	--&	22&	--&	259&	--&	--\\
1-18&	-15.19&	42.98&	-15.54&	42.84&	14&	-9&	412&	68&	0.4\\
1-19&	-15.30&	38.53&	--&	--&	18&	--&	323&	--&	--\\
1-20&	-15.75&	32.80&	-16.27&	32.72&	6&	-7&	308&	82&	0.5\\
1-21&	-15.86&	30.80&	-15.54&	29.47&	10&	13&	324&	347&	1.4\\
1-22&	-16.49&	37.66&	--&	--&	14&	--&	271&	--&	--\\
1-23&	-17.50&	43.82&	--&	--&	12&	--&	426&	--&	--\\
1-24&	-17.67&	21.25&	-17.53&	21.52&	11&	-10&	236&	207&	0.3\\
1-25&	-17.75&	36.51&	--&	--&	11&	--&	410&	--&	--\\
1-26&	-18.48&	35.49&	-18.41&	34.76&	9&	-9&	543&	355&	0.7\\
1-27&	-18.51&	32.72&	--&	--&	8&	--&	445&	--&	--\\
1-28&	-18.59&	23.34&	-17.64&	23.76&	10&	-29&	227&	246&	1.0\\
1-29&	-18.69&	42.70&	--&	--&	14&	--&	409&	--&	--\\
1-30&	-18.76&	30.38&	-18.83&	29.75&	4&	-8&	362&	6&	0.6\\
1-31&	-19.25&	21.21&	--&	--&	12&	--&	304&	--&	--\\
1-32&	-19.99&	25.93&	--&	--&	9&	--&	257&	--&	--\\
1-33&	-20.37&	28.67&	-20.12&	28.45&	8&	-40&	316&	311&	0.3\\
1-34&	-20.58&	35.98&	-20.26&	35.00&	13&	10&	483&	342&	1.0\\
1-35&	-20.61&	42.24&	--&	--&	16&	--&	286&	--&	--\\
1-36&	-20.68&	30.14&	-20.97&	29.30&	8&	-17&	536&	18&	0.9\\
1-37&	-21.00&	32.55&	--&	--&	12&	--&	439&	--&	--\\
1-38&	-21.21&	33.99&	-21.14&	33.95&	12&	-8&	439&	no p.a.&0.1\\
1-39&	-21.49&	37.28&	-21.07&	37.45&	14&	-9&	480&	247&	0.5\\
1-40&	-21.84&	38.64&	-21.84&	39.87&	15&	-7&	401&	180&	1.2\\
1-41&	-21.88&	34.51&	--&	--&	12&	--&	416&	--&	--\\
1-42&	-22.01&	21.74&	--&	--&	7&	--&	223&	--&	--\\
1-43&	-22.23&	24.29&	-21.88&	24.01&	11&	-9&	357&	309&	0.4\\
1-44&	-22.43&	30.24&	--&	--&	10&	--&	389&	--&	--\\
1-45&	-22.50&	18.48&	--&	--&	7&	--&	471&	--&	--\\
1-46&	-22.61&	19.92&	--&	--&	6&	--&	295&	--&	--\\
1-47&	-23.80&	35.21&	-23.91&	35.14&	12&	15&	332&	no p.a.&0.1\\
1-48&	-24.11&	37.56&	-23.94&	36.65&	13&	-8&	329&	349&	0.9\\
1-49&	-24.33&	19.67&	-23.98&	19.57&	10&	-14&	413&	287&	0.4\\
1-50&	-24.57&	21.03&	-24.33&	20.79&	12&	13&	271&	315&	0.3\\
1-51&	-24.82&	28.21&	--&	--&	16&	--&	255&	--&	--\\
1-52&	-25.48&	29.92&	-25.55&	29.19&	14&	9&	282&	5&	0.7\\
1-53&	-25.66&	18.20&	--&	--&	8&	--&	278&	--&	--\\
1-54&	-26.04&	35.53&	-25.76&	35.35&	10&	-31&	349&	302&	0.3\\
1-55&	-26.53&	24.40&	--&	--&	17&	--&	301&	--&	--\\
1-56&	-26.60&	20.79&	-26.28&	20.86&	8&	-5&	270&	257&	0.3\\
1-57&	-26.67&	32.13&	--&	--&	14&	--&	378&	--&	--\\
1-58&	-27.34&	33.32&	-26.74&	33.39&	12&	-20&	394&	263&	0.6\\
1-59&	-27.55&	20.06&	--&	--&	12&	--&	334&	--&	--\\
1-60&	-28.17&	34.33&	-28.49&	33.25&	12&	19&	498&	16&	1.1\\
1-61&	-29.01&	21.52&	-28.49&	21.49&	8&	-32&	293&	274&	0.5\\
1-62&	-29.68&	38.92&	-29.51&	38.81&	19&	-8&	296&	no p.a.&0.2\\
1-63&	-29.68&	23.66&	--&	--&	10&	--&	260&	--&	--\\
1-64&	-30.73&	25.31&	-31.11&	25.66&	15&	8&	257&	132&	0.5\\
1-65&	-30.84&	22.50&	--&	--&	11&	--&	276&	--&	--\\
1-66&	-31.50&	20.75&	--&	--&	13&	--&	335&	--&	--\\
1-67&	-31.89&	37.62&	--&	--&	26&	--&	235&	--&	--\\
1-68&	-31.99&	31.57&	-31.92&	31.11&	17&	-13&	266&	351&	0.5\\
1-69&	-32.24&	25.90&	--&	--&	18&	--&	257&	--&	--\\
1-70&	-32.69&	24.75&	--&	--&	18&	--&	251&	--&	--\\
1-71&	-33.39&	33.74&	--&	--&	28&	--&	241&	--&	--\\
1-72&	-33.78&	27.34&	-34.79&	27.23&	20&	15&	268&	84&	1.0\\
1-73&	-35.67&	39.24&	-35.35&	38.88&	18&	-11&	270&	318&	0.5\\
1-74&	-35.84&	35.07&	--&	--&	27&	--&	438&	--&	--\\
1-75&	-36.22&	28.70&	-36.92&	28.63&	23&	-16&	178&	84&	0.7\\
1-76&	-36.65&	32.97&	-35.74&	32.65&	22&	-16&	228&	289&	1.0\\
1-77&	-37.28&	21.39&	--&	--&	18&	--&	235&	--&	--\\
1-78&	-37.38&	24.29&	--&	--&	22&	--&	236&	--&	--\\
1-79&	-37.45&	32.97&	--&	--&	24&	--&	263&	--&	--\\
1-80&	-37.73&	23.24&	--&	--&	19&	--&	237&	--&	--\\
1-81&	-38.88&	32.94&	-39.94&	32.83&	24&	7&	350&	84&	1.1\\
1-82&	-39.17&	25.83&	--&	--&	22&	--&	296&	--&	--\\
1-83&	-40.60&	22.19&	--&	--&	18&	--&	230&	--&	--\\
1-84&	-40.95&	24.29&	--&	--&	16&	--&	178&	--&	--\\
1-85&	-44.80&	28.94&	--&	--&	21&	--&	179&	--&	--\\
1-86&	-45.29&	35.46&	-45.57&	36.12&	24&	33&	178&	157&	0.7\\
1-87&	-49.17&	26.67&	--&	--&	20&	--&	185&	--&	--\\
\end{longtable}

%
\begin{table*}
\caption{Flows in Peak~2. For an explanation of each column, see
  table~\ref{table_p11}}
\label{table_p2}
\centering
\begin{tabular}{c c c c c c c c c c}
\multicolumn{10}{c}{Flows in Peak~2} \\
\hline \hline
 & \multicolumn{4}{c}{Coordinates of maxima} & velocity of & flow &
maximum & position & displace- \\
No. & \multicolumn{2}{c}{Brightness} & \multicolumn{2}{c}{Velocity} &
surroundings & velocity & brightness & angle & ment \\
\hline 
 & E & N & E & N & km\,s$^{-1}$ & km\,s$^{-1}$ & counts &
degrees & arcsec \\
\hline
 2-1&	19.39&	-6.51&	19.25&	-6.62&	20&	14&	261&	no p.a.&0.2\\
 2-2&	19.00&	 9.45&	19.42&	 9.41&	34&	9&	233&	275&	0.4\\
 2-3&	18.59&	-5.43&	18.73&	-5.36&	20&	9&	508&	no p.a.&0.2\\
 2-4&	18.38&	 3.67&	17.75&	 3.71&	34&	-12&	532&	93&	0.6\\
 2-5&	17.64&	-0.74&	17.40&	-1.23&	24&	-7&	391&	27&	0.5\\
 2-6&	16.76&	 8.72&	16.84&	 8.72&	41&	7&	407&	no p.a.&0.1\\
 2-7&	16.34&	 9.73&	--&	--&	41&	--&	391&	--&	--\\
 2-8&	16.20&	-6.47&	16.10&	-6.51&	29&	10&	357&	no p.a.&0.1\\
 2-9&	16.17&	-0.49&	16.66&	-0.67&	24&	-9&	331&	290&	0.5\\
2-10&	15.82&	-1.61&	15.47&	-1.12&	24&	10&	860&	144&	0.6\\
2-11&	14.81&	 6.58&	--&	--&	34&	--&	287&	--&	--\\
2-12&	14.18&	 9.80&	13.90&	 9.52&	40&	-8&	210&	45&	0.4\\
2-13&	13.48&	-1.75&	--&	--&	30&	--&	475&	--&	--\\
2-14&	13.37&	-0.46&	13.51&	-0.42&	30&	7&	506&	no p.a.&0.1\\
2-15&	12.91&	 8.26&	--&	--&	35&	--&	209&	--&	--\\
2-16&	12.36&	 0.67&	--&	--&	32&	--&	454&	--&	--\\
2-17&	12.22&	-2.66&	12.36&	-3.12&	25&	-6&	244&	343&	0.5\\
2-18&	12.15&	 2.28&	--&	--&	32&	--&	436&	--&	--\\
2-19&	11.65&	 3.95&	11.62&	 3.88&	32&	-8&	404&	no p.a.&0.1\\
2-20&	10.81&	 1.23&	10.57&	 1.64&	32&	11&	456&	150&	0.5\\
2-21&	10.53&	-0.94&	10.43&	-0.31&	32&	8&	361&	171&	0.6\\
2-22&	10.40&	 8.75&	10.81&	 8.78&	34&	-7&	188&	265&	0.4\\
2-23&	10.11&	-6.51&	--&	--&	23&	--&	273&	--&	--\\
2-24&	 9.66&	 1.19&	 9.84&	 1.01&	31&	7&	349&	no p.a.&0.2\\
2-25&	 9.56&	 2.17&	 9.38&	 3.01&	31&	-10&	341&	168&	0.9\\
2-26&	 8.86&	 1.78&	 8.78&	 1.78&	31&	-12&	370&	no p.a.&0.1\\
2-27&	 7.70&	 6.30&	 8.12&	 6.20&	30&	-7&	220&	284&	0.4\\
2-28&	 7.53&	-3.64&	--&	--&	27&	--&	308&	--&	--\\
2-29&	 7.07&	-1.05&	 7.03&	-1.01&	30&	-6&	335&	no p.a.&0.0\\
2-30&	 6.82&	-4.69&	--&	--&	26&	--&	308&	--&	--\\
2-31&	 6.62&	 6.86&	 6.51&	 6.65&	26&	-11&	182&	no p.a.&0.2\\
2-32&	 6.62&	 2.62&	 6.72&	 2.73&	30&	32&	283&	no p.a.&0.1\\
2-33&	 6.51&	-3.05&	 6.47&	-3.01&	28&	-6&	350&	no p.a.&0.0\\
2-34&	 6.30&	 1.09&	 6.86&	 1.33&	30&	26&	310&	246&	0.6\\
2-35&	 6.30&	-0.56&	 6.30&	-0.42&	30&	15&	498&	no p.a.&0.1\\
2-36&	 5.95&	 7.18&	 5.99&	 7.24&	26&	-13&	194&	no p.a.&0.1\\
2-37&	 5.25&	-3.33&	 5.14&	-3.05&	32&	10&	530&	159&	0.3\\
2-38&	 5.07&	-2.77&	 5.07&	-2.24&	30&	9&	323&	180&	0.5\\
2-39&	 4.80&	-8.23&	 4.24&	-7.98&	20&	-7&	205&	114&	0.6\\
2-40&	 4.62&	10.85&	 4.41&	10.95&	20&	-18&	531&	no p.a.&0.2\\
2-41&	 3.82&	-2.38&	 3.64&	-2.56&	31&	11&	209&	no p.a.&0.2\\
2-42&	 3.12&	 7.66&	 3.01&	 7.45&	20&	-8&	305&	no p.a.&0.2\\
2-43&	 2.62&	 0.94&	 2.66&	 1.09&	20&	-7&	232&	no p.a.&0.1\\
2-44&	 1.92&	 1.54&	 1.68&	 1.58&	20&	-14&	227&	no p.a.&0.2\\
2-45&	 1.58&	 5.28&	 1.78&	 4.93&	18&	6&	542&	329&	0.4\\
2-46&	 1.37&	 4.03&	 1.92&	 3.67&	15&	4&	377&	302&	0.7\\
2-47&	 0.74&	 6.05&	 0.77&	 6.41&	10&	-13&	372&	186&	0.4\\
2-48&	 0.14&	 4.03&	-0.42&	 4.13&	10&	-8&	316&	101&	0.6\\
2-49&	-0.07&	 8.05&	-0.14&	 8.30&	12&	-15&	229&	164&	0.3\\
2-50&	-1.05&	 9.80&	--&	--&	16&	--&	192&	--&	--\\
\hline \hline
\end{tabular}
\end{table*}
%
\begin{table*}
\caption{Flows in Region B. For an
  explanation of each column, see table \ref{table_p11}} 
\label{table_cz}
\centering
\begin{tabular}{c c c c c c c c c c}
\multicolumn{10}{c}{Flows in Region B} \\
\hline \hline

 & \multicolumn{4}{c}{Coordinates of maxima} & velocity of & flow & maximum &
position & displace- \\
No. & \multicolumn{2}{c}{Brightness} & \multicolumn{2}{c}{Velocity} &
surroundings & velocity & brightness & angle & ment \\

\hline 
 & E & N & E & N & km\,s$^{-1}$ & km\,s$^{-1}$ & counts &
degrees & arcsec \\
\hline
 B-1&	-2.20&	5.60&	-2.13&	5.46&	4&	-14&	240&	no p.a.&0.2\\
 B-2&	-3.26&	-12.43&	-3.61&	-12.36&	22&	-9&	174&	101&	0.4\\
 B-3&	-3.99&	6.12&	-3.61&	6.72&	5&	-22&	348&	213&	0.7\\
 B-4&	-4.41&	-6.05&	-4.83&	-5.70&	22&	-8&	192&	130&	0.5\\
 B-5&	-4.66&	5.70&	-5.57&	5.64&	5&	-34&	432&	86&	0.9\\
 B-6&	-4.93&	-8.89&	-5.04&	-8.89&	22&	-7&	145&	no p.a.&0.1\\
 B-7&	-7.77&	7.80&	-7.95&	7.84&	7&	-28&	208&	no p.a.&0.2\\
 B-8&	-8.05&	13.90&	-7.98&	13.69&	14&	-24&	255&	no p.a.&0.2\\
 B-9&	-8.05&	7.66&	-7.24&	7.32&	7&	-20&	199&	293&	0.9\\
B-10&	-8.19&	7.32&	-7.53&	7.00&	7&	-24&	198&	295&	0.7\\
B-11&	-8.36&	10.29&	-8.26&	10.36&	12&	-12&	175&	no p.a.&0.1\\
B-12&	-8.68&	14.77&	-8.78&	14.56&	14&	-21&	365&	no p.a.&0.2\\
B-13&	-8.93&	13.61&	-9.17&	13.12&	14&	-10&	343&	27&	0.5\\
B-14&	-9.69&	-8.75&	-9.80&	-8.72&	27&	-13&	181&	no p.a.&0.1\\
B-15&	-9.90&	-8.19&	-9.98&	-8.05&	27&	-14&	182&	no p.a.&0.2\\
B-16&	-10.64&	-0.49&	-10.71&	-0.07&	13&	-13&	261&	171&	0.4\\
B-17&	-10.99&	-16.94&	--&	--&	26&	--&	231&	--&	--\\
B-18&	-11.80&	2.13&	-11.94&	2.38&	7&	-13&	362&	150&	0.3\\
B-19&	-11.94&	1.58&	-11.94&	1.54&	7&	-16&	392&	no p.a.&0.0\\
B-20&	-12.32&	1.50&	-12.49&	1.58&	7&	-18&	368&	no p.a.&0.2\\
B-21&	-12.81&	2.94&	-12.64&	2.97&	7&	-14&	365&	no p.a.&0.2\\
B-22&	-13.12&	1.92&	-13.23&	2.00&	7&	-17&	307&	no p.a.&0.1\\
B-23&	-13.69&	-0.52&	-13.30&	0.07&	8&	-20&	531&	213$^\star$&	0.7\\
B-24&	-13.76&	11.44&	-13.27&	11.86&	3&	-16&	226&	229&	0.6\\
B-25&	-14.35&	11.90&	-13.82&	12.01&	3&	-23&	212&	259&	0.5\\
B-26&	-14.73&	-7.56&	-14.45&	-6.89&	20&	-7&	178&	203$^\star$&	0.7\\
B-27&	-15.36&	-11.90&	-15.36&	-11.62&	10&	-12&	200&	180&	0.3\\
B-28&	-15.89&	12.95&	-14.77&	13.06&	3&	-20&	216&	265&	1.1\\
B-29&	-15.99&	7.28&	-15.82&	7.00&	7&	-20&	163&	328&	0.3\\
B-30&	-16.62&	13.93&	-16.66&	13.90&	3&	-11&	167&	no p.a.&0.0\\
B-31&	-16.73&	5.99&	--&	--&	7&	--&	221&	--&	--\\
B-32&	-17.29&	8.68&	-17.15&	8.36&	8&	-20&	190&	336&	0.3\\
B-33&	-17.57&	10.05&	-17.57&	9.98&	4&	-17&	400&	no p.a.&0.1\\
B-34&	-17.60&	7.80&	-17.60&	7.95&	8&	-20&	206&	no p.a.&0.1\\
B-35&	-17.78&	0.17&	-17.40&	-0.38&	10&	-26&	358&	325&	0.7\\
B-36&	-17.95&	-1.15&	-17.82&	-2.13&	10&	-37&	404&	352&	1.0\\
B-37&	-18.48&	0.00&	-17.85&	0.38&	10&	-18&	363&	239$^\star$&	0.7\\
B-38&	-18.73&	2.52&	-18.51&	3.05&	15&	-14&	234&	202&	0.6\\
B-39&	-18.76&	5.46&	--&	--&	10&	--&	239&	--&	--\\
B-40&	-18.90&	7.95&	--&	--&	6&	--&	281&	--&	--\\
B-41&	-19.15&	6.65&	-19.08&	6.72&	8&	-19&	301&	no p.a.&0.1\\
B-42&	-19.67&	16.24&	-19.84&	16.91&	7&	-19&	401&	165$^\star$&	0.7\\
B-43&	-20.44&	-6.76&	-20.23&	-6.51&	7&	-36&	291&	221$^\star$&	0.3\\
B-44&	-22.78&	7.28&	-22.78&	7.32&	10&	-39&	173&	no p.a.&0.0\\
B-45&	-23.10&	6.05&	-22.54&	6.12&	10&	-15&	208&	263&	0.6\\
B-46&	-23.91&	6.34&	-23.91&	6.34&	10&	-29&	293&	no p.a.&0.0\\
B-47&	-24.75&	0.88&	-24.85&	0.98&	8&	-20&	256&	no p.a.&0.1\\
B-48&	-25.48&	16.34&	-25.02&	16.45&	5&	-14&	356&	257&	0.5\\
B-49&	-25.83&	0.35&	-25.66&	0.46&	7&	-7&	188&	no p.a.&0.2\\
B-50&	-27.26&	16.49&	-27.34&	16.45&	5&	-15&	305&	no p.a.&0.1\\
B-51&	-27.55&	11.65&	-26.99&	12.01&	12&	-19&	273&	238&	0.7\\
B-52&	-30.03&	-8.40&	--&	--&	18&	--&	379&	--&	--\\
B-53&	-30.91&	-10.08&	--&	--&	18&	--&	368&	--&	--\\
B-54&	-31.26&	5.95&	-30.94&	6.02&	17&	-7&	314&	257&	0.3\\
B-55&	-33.28&	7.18&	--&	--&	18&	--&	249&	--&	--\\
B-56&	-33.92&	16.17&	-33.35&	16.10&	12&	-9&	202&	277&	0.6\\
\hline \hline
\end{tabular}
\end{table*}

\end{document}